\documentclass{agujournal}

\journalname{JGR-Space Physics}

\begin{document}

%
%


\title{Using the Coronal Evolution to Successfully Forward Model CMEs' In Situ Magnetic Profiles}

%
%




\authors{C. Kay\affil{1} and N. Gopalswamy\affil{1}}


\affiliation{1}{Solar Physics Laboratory, NASA Goddard Space Flight Center, Greenbelt, MD, USA}




\correspondingauthor{C. Kay}{christina.d.kay@nasa.gov}




\begin{keypoints}
\item ForeCAT simulations reproduce the deflection and rotation of 45 CMEs observed by STEREO 
\item FIDO simulations, driven by ForeCAT results, reproduce the in situ magnetic field observed by ACE
\item ForeCAT-driven results match the in situ data better than results driven by coronal reconstructions 
\end{keypoints}

%
%


\begin{abstract}
Predicting the effects of a coronal mass ejection (CME) impact requires knowing if impact will occur, which part of the CME impacts, and its magnetic properties.  We explore the relation between CME deflections and rotations, which change the position and orientation of a CME, and the resulting magnetic profiles at 1 AU.  For 45 STEREO-era, Earth-impacting CMEs, we determine the solar source of each CME, reconstruct its coronal position and orientation, and perform a ForeCAT \citep{Kay15} simulation of the coronal deflection and rotation.  From the reconstructed and modeled CME deflections and rotations we determine the solar cycle variation and correlations with CME properties. We assume no evolution between the outer corona and 1 AU and use the ForeCAT results to drive the FIDO in situ magnetic field model \citep{Kay17FIDO}, allowing for comparisons with ACE and Wind observations. We do not attempt to reproduce the arrival time.  On average FIDO reproduces the in situ magnetic field for each vector component with an error equivalent to 35\% of the average total magnetic field strength when the total modeled magnetic field is scaled to match the average observed value.  Random walk best fits distinguish between ForeCAT's ability to determine FIDO's input parameters and the limitations of the simple flux rope model. These best fits reduce the average error to 30\%. The FIDO results are sensitive to changes of order a degree in the CME latitude, longitude, and tilt, suggesting that accurate space weather predictions require accurate measurements of a CME's position and orientation.   
\end{abstract}

%
%

%


%
%
%
%

\section{Introduction}
Coronal mass ejections (CMEs) can drive intense magnetic storms at Earth and throughout the rest of the solar system.  Observations at Mars from the Mars Atmosphere and Volatile Evolution (MAVEN) spacecraft \citep{Jak15} show the response of the Martian magnetosphere and atmosphere to a CME impact, and suggest that CME impacts may have played a prominent role in the evolution of the Martian atmosphere.  Predicting the effect that a CME may have on a planetary environment requires understanding the path that a CME takes as it propagates away from the Sun and the properties of the CME itself. 

The earliest spaceborne coronagraph measurements showed that the position angle of a CME often changes as it propagated through the coronal field-of-view.  \citet{Hil77} found a trend of CMEs deflecting toward the equator in the Skylab coronagraph observations.  In the Solar Maximum Mission observations this trend was no longer present \citep{Mac86}.  CMEs still underwent deflections, but deflections occurred both toward and away from the equator.  Changes in position angle typically correspond to latitudinal deflections, but via geometric reconstruction techniques and the multiple viewpoints allowed by the twin Solar Terrestrial Relations Observatory (STEREO) spacecraft, longitudinal deflections have also been observed \citep[e.g.][]{Gui11,Isa14,Mos15}.

CME deflections are frequently attributed the magnetic forces, which tend to be the dominant forces in the low corona.  On global scales, CMEs deflect toward the Heliospheric Current Sheet (HCS) and away from coronal holes \citep[e.g.][]{Fil01,Cre04, Gop09, Kil09}, following the background gradients in the magnetic energy.  This motion can explain the discrepancy between the trends observed in the Skylab and Solar Maximum Mission observations.  The Skylab observations occurred during solar minimum, when latitudinal deflections toward the equator would be expected due to the flat HCS, whereas during solar maximum the HCS is more inclined and more variety in the direction of deflection would be expected.

Similar to deflection changing a CME's latitude or longitude, rotation causes the tilt of a CME, typically measured counterclockwise with respect to the solar equator, to change from its initial value.  Rotation is often difficult to measure in the low corona where the CME is simultaneously deflecting, rotating, and expanding nonuniformally \citep{Sav10, Nie13} but a rotation has been inferred from some observations \citep{Vou11,Tho12,Isa14}.  Simulations of CMEs often show rotation without any ambiguity \citep[e.g.][]{Lyn09,Kli12}, which often occur as a result of the kink instability so that the flux rope handedness and the direction of rotation are intrinsically coupled.

Observations and simulations show that the largest CME deflections and rotations occur in the corona \citep{Byr10,Isa14,Kay15,Kay17AR}, but observations suggest that interplanetary deflections may also occur \citep[e.g][]{Wan04, Liu10a, Lug10, Wan14}.  Multiple interplanetary coronal mass ejections (ICMEs) interacting can lead to deflections \citep[e.g.][]{Xio07, Lug12, Lug13}.  \citet{Wan04} suggest that an isolated ICME deflects longitudinally to follow the Parker spiral so that fast ICMEs are forced eastward due to solar wind piling up ahead of the ICME and slow ICMEs deflect westward due the accumulation behind.  \citet{Kay15} and \citet{Kay15AM}, however, suggest that the magnetic forces beyond the low corona are not strong enough to yield significant deflections.  \citet{Sac17} study the magnetic forces driving radial acceleration, which originate from the same magnetic background as the tangential deflection forces, and find that these forces can become negligible by 4 $R_{\odot}$ for fast CMEs but can still influence the radial velocity as far as 12 to 50 $R_{\odot}$ for slow CMEs.  

At Earth, geomagnetic disturbances are often measured by the disturbance storm time (Dst) index, a measure of the Earth's ring current.  Observations show that the intensity of the storm (as measured by Dst) tends to be correlated with the magnitude of an ICME's southward magnetic field \citep{Yur05} or the product of the ICME velocity and the southward magnetic field \citep{Gop08}.  Space weather predictions therefore depend critically on accurately determining the magnetic field of ICMEs, which depend not only on the magnetic field of the erupting flux rope, but its final orientation including any deflection or rotation before impact.

Many of the studies of ICME in situ magnetic field focus on inferring information about the ICME from the magnetic profile rather than predicting the magnetic profile from known ICME properties.  \citet{Hu02} introduce a technique for fitting the two-dimensional magnetostatic Grad-Shafranov equation to an in situ profile from a single spacecraft.  This technique has been applied extensively to a large number of observed ICMEs \citep[e.g.][]{Mos08,Mos09,AlH11}.   \citet{Lep90} introduce a technique for fitting the Lundquist solution for a force free flux rope to an in situ profile in order to determine the orientation of the flux rope.  This model has been applied to all the ICMEs observed by Wind \citep[e.g.][]{Lep11,Lep15}.

Few studies thus far have used remote coronagraph or heliospheric imager observations to forward model the full in situ magnetic profile of an ICME (e.g. one case in \citet{Kun10} and two cases in \citet{Isa16}) with the eight cases of \citet{Sav15} representing the largest study of this kind.  \citet{Gop17b} combine geometrical properties from the corona with the observed reconnection flux to predict the magnetic flux at 1 AU for over 50 CMEs, but do not compare the full magnetic profiles.  Studies that combine remote and in situ measurements for a large number of CMEs ($>$ 10) tend to have another focus for the in situ data, such as predicting impacts and misses \citep{Rod11} or the arrival time \citep{Mos14}, or determining the shape of the CME \citep[e.g][]{Owe08, Jan13,Jan15} or other near-Earth characteristics \citep[e.g.][]{Kil09, Ber11, Tem17}.  Magnetohydrodynamic (MHD) simulations of ICMEs are often compared with observations, yielding reasonable results \citep[e.g.][]{Zho14,Shi16,Jin17}, but these simulations typically do not run on time scales that would be reasonable for future space weather predictions or ensemble studies.

\citet{Isa13} and \citet{Isa14} study the deflection and rotation of CMEs by combining both coronagraph observations and in situ magnetic field observations.  For 15 CMEs, the coronal position is reconstructed and the position and orientation near 1 AU inferred from the Grad-Shafranov reconstruction combined with geometric arguments and an assumed longitudinal deflection along the Parker Spiral according to \citet{Wan04}.  The Parker-spiral longitudinal deflection corresponds to small values, typically less than 5$^{\circ}$, as the CME speed does not differ greatly from the background solar wind CME for the \citet{Isa13,Isa14} CMEs.  This small range of interplanetary longitudinal deflections can result in large inferred interplanetary latitudinal deflections with values reaching as large as 35$^{\circ}$ and most cases falling between 10$^{\circ}$ and 20$^{\circ}$.  

In a recent study, \citet{Woo17} determine the full trajectory of ICMEs observed by Wind.  For 28 cases, \citet{Woo17} use STEREO observations to follow the corresponding CME from the Sun to near 1 AU.  \citet{Woo17} find that the CME position, orientation, and size inferred from their coronagraph or heliospheric imager reconstructions often do not match the values determined from the in situ magnetic field.

In this paper we combine observations and simulations to present the largest-to-date study of the coronal deflections and rotations of CMEs and the resulting effects on the in situ magnetic profiles near Earth.  In section \ref{ObsMeth} we describe the data set and the method by which the coronagraph observations are processed.  In section \ref{models} we describe ForeCAT, which simulates the coronal deflection and rotation of the CME, and FIDO, which takes the ForeCAT results and produces an in situ profile of the CME's magnetic field.  Section \ref{Obs} contains the coronal results from both the observations and simulations and section \ref{1AU} contains the in situ results near 1 AU.

\section{Data Set and Coronal Observations}\label{ObsMeth}
To select the CMEs in this paper, we begin with the \citet{Ric10} list of near-Earth ICMEs (updated list at http://www.srl.caltech.edu/ACE/ASC/DATA/level3/icmetable2.htm, hereafter referred to as the RC ICME list).  From these, we select only the ICMEs that have an identified LASCO coronal CME counterpart within the list.  We emphasize that we fully rely on the RC identifications and do not study the evolution of these CMEs between the corona and 1 AU.  It can be difficult to unambiguously associate in situ CMEs with their coronal counterparts if multiple CMEs occur in rapid succession, especially if CME-CME interactions occur.  However, many of these CMEs appear in other lists (such as in \citet{Gop13} and \citet{Sel16}) and we find that they have consistent identification for the coronal counterparts of the ICMEs.  Finally, we exclude any CMEs that do not have at least COR2 coronagraph observations from both STEREO spacecraft, yielding a final set of 45 CMEs between November 2007 and June 2014, occurring between the late declining phase of Solar Cycle 23 and the beginning of the maximum of Solar Cycle 24.  This set includes ICMEs that satisfy the criteria to qualify as magnetic clouds (MCs, \citet{Bur81}) and more complicated or less clear structures that would be classified as non-MCs or ejecta.  We make no distinction between the two sets within this paper, showing that FIDO can reproduce both types, though we compare our results with the ICME ``quality'' in Section \ref{1AU}, and the highest quality cases typically correspond to MCs.  The ICMEs classified as MCs in the RC ICME lists have an ``MC'' by the quality in Table 2 of the supplementary material.

We determine an approximate initial location using observations from a combination of extreme ultraviolet (EUV) imagers to determine whether each CME erupts from an active region and, if so, which part of the active region erupts.  The Solar and Heliospheric Observatory's Extreme Ultraviolet Imaging Telescope (SOHO/EIT 195 \AA) and the Solar Dynamics Observatory Atmospheric Imaging Assembly (SOHO/AIA 171 \AA{} or 193 \AA) tend to provide a face on view of the post-eruption arcade and the STEREO Extreme Ultraviolet Imager (EUVI 195 \AA) can show the location of filaments off the solar limb for many of these Earth-directed CMEs.  We combine this approximate location with an SDO Helioseismic and Magnetic Imager (HMI) magnetogram (or SOHO Michelson Doppler Imager (MDI) magnetogram for CMEs before May 2010) to determine the polarity inversion line from which each CME erupts, yielding a better approximation to the initial latitude, longitude, and orientation of each CME.  For the cases that also appear in \citet{Gop13} and \citet{Sel16}, we find good agreement in the initial latitude and longitude of the CMEs.

We use the Graduated Cylindrical Shell model \citep[GCS,][]{The06} to reconstruct the position and orientation of each CME at farther distances.  Simultaneously for both the STEREO A and B viewpoints, we visually fit the GCS shape to running difference images from both COR1 and COR2.  We obtain a single fit for each of the COR1 and COR2 field-of-views rather than reconstructing the full trajectory at as many time possible due to the large sample size of 45 CMEs.  Table 1 of the supplementary material lists the results for the free parameters of the GCS reconstructions for both COR1 and COR2.  Each CME is given an identification number (ID) and the listed times correspond to the first appearance in the COR1 field-of-view.  Several CMEs cannot be observed in COR1 and the COR2 times are listed, which is indicated with an $^{*}$. Table 1 includes the CME latitude, longitude, tilt (measured counterclockwise from the solar equator), radial distance of the front of the CME, $R$, aspect ratio, $\kappa$, and angular width.

Near 1 AU we consider in situ data from both ACE and Wind.  Many of the ICMEs within the RC ICME list are also identified in the Wind ICME list (available at http://wind.nasa.gov/ICMEindex.php).  Both lists contain start and stop times for each CME, and these two independent measurements are often not in agreement despite the magnetic field not varying significantly between spacecraft.  The stop times of the CMEs tend to vary the most as it is often difficult to define a precise boundary for the back of the CME.  When comparing with the in situ data, we tend to use the most conservative set of boundaries, which may combine times from the two different lists.

\section{ForeCAT and FIDO}\label{models}
To simulate each CME's coronal deflection and rotation, we use Forecasting a CME's Altered Trajectory \citep[ForeCAT,][]{Kay13, Kay15}.  ForeCAT calculates the magnetic tension and magnetic pressure gradients from an HMI- or MDI-driven Potential Field Source Surface (PFSS) model of the background solar magnetic field, the net force determining the deflection and the torque creating a rotation about an axis running through the nose of the CME to the center of the Sun.  ForeCAT uses simple analytic or empirical models to describe a CME's radial velocity and angular width as a function of radial distance and the deflection and rotation cause a change in the CME's latitude, longitude, and tilt on top of the propagation and expansion.  ForeCAT CMEs can be simulated to any radial distance, but, as seen in \citet{Kay15AM}, the deflection and rotation become negligible by about 10 $R_{\odot}$ because the magnetic forces rapidly decrease with radial distance. 

Within ForeCAT, the CME flux rope is represented with a rigid, undeformable torus. The shape is specified by the ratio of the height to width, $A$, and the ratio of the cross-sectional width to the width, $B$ (see Figure 2 of \citet{Kay15}, $A=a/c$, $B=b/c$).  With the current rigid shape the torus will increase in size as the CME expands but the shape parameters $A$ and $B$ do not change during the simulation.  The front of the torus is represented numerically with a grid consisting of 15 points in the toroidal direction and 13 points in the poloidal direction.  ForeCAT calculates the forces at each grid point simultaneously via parallelization on a Graphics Processing Unit (GPU).  Accordingly, ForeCAT runs very efficiently, taking approximately 10 seconds to simulate a CME to 10 $R_{\odot}$ on an average desktop computer.

For this paper, we simulate CMEs out to a distance of 50 $R_{\odot}$ with ForeCAT.  Any simulated deflection or rotation due to magnetic forces is negligible beyond this distance.  The observed PIL location puts constraints on the initial latitude, longitude, and tilt of the ForeCAT CME, and the observed velocity and angular width constrain the propagation and expansion models.  The CMEs follow a three-phase propagation model, similar to that of \citet{Zha06}.  The CME begins at some initial velocity, $v_{min}$, then linearly accelerates to a coronal speed, $v_{cor}$, between two radial distances.  These radial distances and $v_{min}$ are relatively unconstrained from the observations due to only reconstructing each CME at a maximum of two different distances.  The simulated CMEs expand exponentially from some minimum to a maximum angular width.  The angular width reconstructed from COR2 determines the maximum angular width, and the minimum can be constrained by the value from COR1.  We also include a linearly increasing CME mass below 10 $R_{\odot}$ to mimic the effects of solar wind swept up by the erupting CME.  The final mass is determined from the LASCO CME catalog values, where possible, and the initial mass is set at half the final value.  The shape parameters are entirely unconstrained, but we restrict them to ``reasonable'' values - A between 0.5 and 1 and B between 0.1 and 0.3.

We couple the ForeCAT results to the ForeCAT In situ Data Observer \citep[FIDO,][]{Kay17FIDO}.  FIDO uses the CME position and tilt from ForeCAT to orient a CME flux rope.  FIDO propagates the flux rope past a synthetic spacecraft, determining the spacecraft's distance from the toroidal axis and the orientation of the CME's toroidal and poloidal directions as a function of time.  The distance from the axis determines the magnitude of the toroidal and poloidal magnetic field and the unit vectors from the CME orientation allow for conversion to Geocentric Solar Ecliptic (GSE) coordinates and direct comparison to the ACE and Wind data.  Note that we do not simulate the propagation between 50 $R_{\odot}$ and 1 AU.  We assume no significant changes in position and orientation occur between these distances and make no attempt to simulate the arrival time of each ICME.

\citet{Kay17FIDO} used the Lundquist solution for toroidal and poloidal components, $B_{t,FF}$ and $B_{p,FF}$, of a force free flux rope

\begin{linenomath*}
\begin{equation}\label{eq:BtFF}
B_{t,FF} = B_0 J_0\left(\frac{2.4d}{b}\right)
\end{equation}
\end{linenomath*}

\begin{linenomath*}
\begin{equation}\label{eq:BpFF}
B_{p,FF} = B_0 H J_1\left(\frac{2.4d}{b}\right)
\end{equation}
\end{linenomath*}

where $B_0$ is the axial magnetic field strength, $H$ the handedness of the flux rope ($=\pm$1), $d$ the impact parameter, $b$ the cross-sectional radius of the flux rope, and $J_0$ and $J_1$ are Bessel functions.  The value 2.4 causes $B_{t,FF}$ to reach zero at the edge of the flux rope ($d=b$).  We note that this simplification assumes that the flux rope can locally be described as cylindrical at any time step, the full solution for a Lundquist type toroidal flux rope can be found in \citet{Ber13}.  In this paper we also consider a circular flux rope model \citep{Nie16} 

\begin{linenomath*}
\begin{equation}\label{eq:Btcirc}
B_{t,circ} = B_0 \left(\tau - \frac{d}{b}\right)^{(n+1)}
\end{equation}
\end{linenomath*}

\begin{linenomath*}
\begin{equation}\label{eq:Bpcirc}
B_{p,circ} = \frac{B_0}{C_1} H \left(\frac{n+1}{m+2}\right)\left(\frac{d}{b}\right)^{(m+1)}
\end{equation}
\end{linenomath*}

where $m$, $n$, $\tau$, and $C_1$ are constants that parameterize the model.  Here we assume values of zero and one for $m$ and $n$. $\tau$, which determines the profile of the toroidal field, and $C_1$, which determines the ratio of the toroidal and poloidal components, are left as free parameters but assumed to be between 0.5 and 2.

The ForeCAT results determine the latitude, longitude, and tilt of the FIDO flux rope, and the speed at 1 AU from the RC ICME list determines the duration of the passage, which begins at the start time chosen from the RC and Wind ICME lists.  We use the same angular width as at the end of the ForeCAT simulation and assume that the flux rope keeps expanding self-similarly, maintaining a fixed angular width as it passes over the synthetic spacecraft.  We allow for changes in the shape ratios $A$ and $B$ between ForeCAT and FIDO to mimic an evolution in the flux rope shape during interplanetary propagation.  Most often we only increase $B$ to allow for an increase in the cross section relative to the total width.  

The helicity or handedness of the flux rope and the sign of axial field $B_0$ can be inferred from the HMI/MDI magnetogram.  \citet{Bot98} determine the helicity from the hemisphere of the eruption (negative in the north, positive in the south), and determine the direction of the toroidal field from the polarity of the leading and trailing flux systems of the AR.  This simple method works best for CMEs with relatively low inclination that initiate from isolated ARs with little non-dipolar structure, but may not be appropriate for more inclined CMEs erupting from more complicated ARs.  The Bothmer-Schwenn relation also breaks down for ARs that have the opposite helicity as expected for their hemisphere, which may include as much as 40\% of ARs \citep{Pet14}. \citet{Pal17} perform a detailed study of the source region of two CMEs showing that a CME's helicity and axial orientation can be estimated best by using a combination of proxies observable in the corona.  In this work we attempt to identify the flux rope foot points from the EUV observations and compare with the polarities in the magnetogram to obtain an estimate of the expected handedness and axial field direction. 

The final free parameter is the magnitude of $B_0$.  \citet{Kay17FIDO} kept this as a free parameter, determining the appropriate value for each case.  Here, we take the average of the total magnetic field strength during the middle four hours of both the ACE ICME and model flux rope (with $B_0$ initially unassigned, or set to one) and use the ratio of the two averages to determine the value of $B_0$.  This reduces the number of free parameters, which helps with finding the best fits in section \ref{BestFits}, and we find that this normalization tends to yield a reasonable representation of the total magnetic field strength over the full duration.

We note that the FIDO model follows nearly the same algorithm as the in situ prediction model of \citet{Sav15} but has several key differences.  First, \citet{Sav15} use a GCS reconstruction to orient their flux rope.  This reconstruction technique has large uncertainties due to the imprecise nature of visual fits to white-light coronagraph images and the frequent occurrence of degeneracies between the model parameters.  Typically uncertainties of 5$^{\circ}$ in latitude and 10$^{\circ}$ in longitude and tilt are assumed \citep{The09}.  \citet{Kay17FIDO} shows that this level of uncertainty in the position and orientation can lead to a wide range of in situ results and suggest instead using the results of a simulation such as ForeCAT, which should have less uncertainty on average.  \citet{Sav15} also do not include any expansion in their flux rope as it passes the spacecraft so that the profile of the total magnetic field strength is symmetric in time.  CME expansion creates a long, decreasing tail in the observed total magnetic field strength \citep[e.g][]{Far93}, and \citet{Kay17FIDO} show that including expansion tends to match to the observed tails.

\section{Example Case}\label{Ex}
\begin{figure}[h!]
\centering
\includegraphics[width=4in]{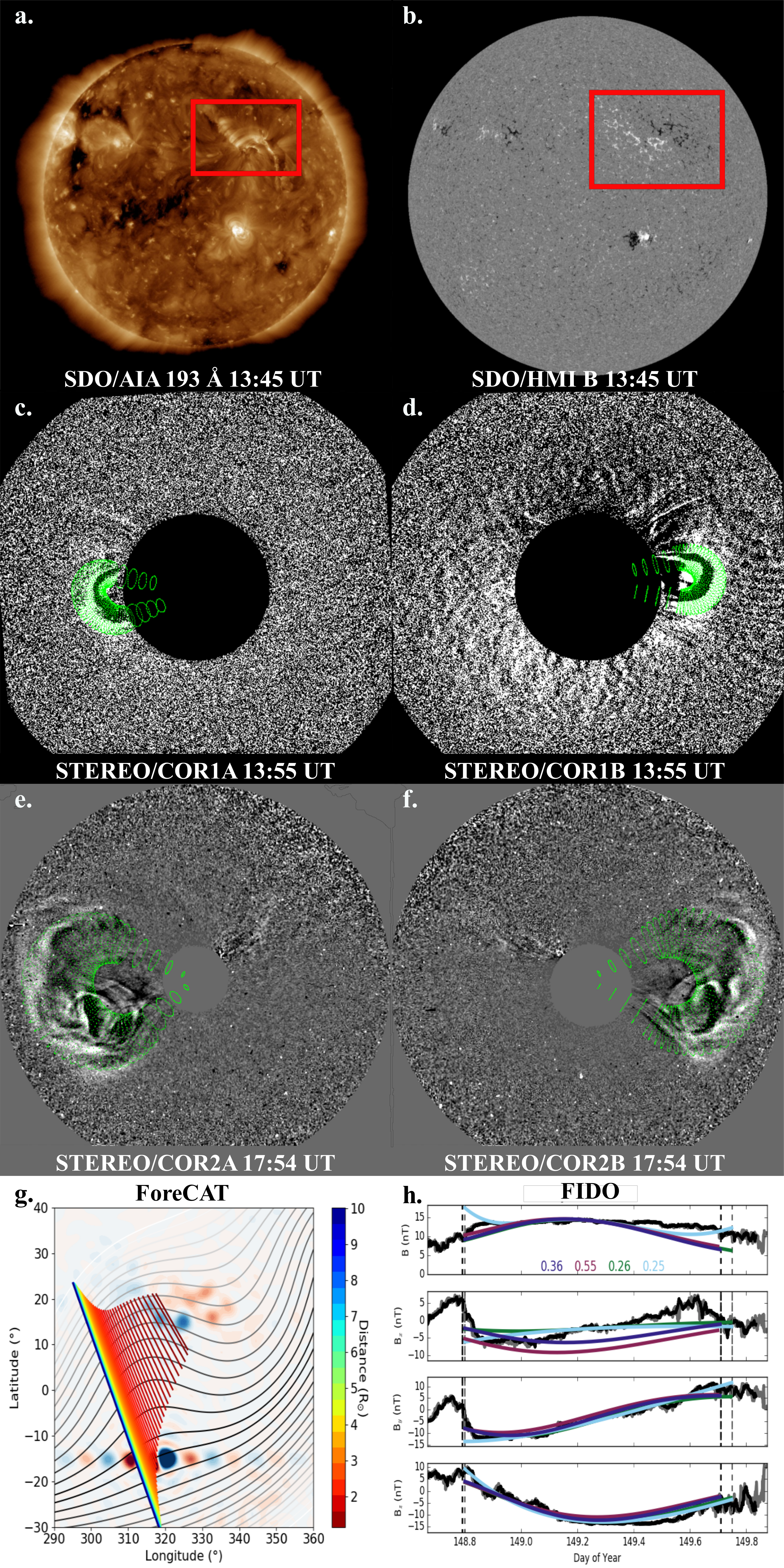}
\caption{Depiction of the step-by-step analysis for a single case (CME 7 in the tables).  All SDO and STEREO times are on 24 May 2010.}
\label{fig:ex}
\end{figure}

Before describing the results of each step in detail, we briefly describe the full process for a single case to provide a better overview.  Figure \ref{fig:ex} shows the step-by-step process for a single CME (CME 7, which erupted on 24 May 2010.)  We start with the the paired coronal and in situ times from the Richardson and Cane ICME list and begin by looking for the site of the eruption on the solar disk in the hours preceding the coronal time.  Figure \ref{fig:ex}(a) shows an SDO/AIA 193 \AA{} image from 24 May 2010 at 13:45 UT, the red box indicates the location of the arcade formed by reconnection behind the CME.  We compare this location with an HMI magnetogram (Figure \ref{fig:ex}(b), red box corresponds roughly to the same region as in Figure \ref{fig:ex}(a)) and determine the location of the erupting PIL.  We then visually fit the GCS model to STEREO images from both COR1 and COR2 (Figure \ref{fig:ex}(c)-(f)).

We then use the observational results with our simulations.  Some of the information (such as CME width, speed, and initial location) are used as inputs for the ForeCAT model, but the model results are compared with the reconstructed coronal position and orientation.  Figure \ref{fig:ex}(g) shows the results of the ForeCAT simulation for this CME.  The background shows the PFSS magnetic field strength at the surface (color contours) and at the source surface (line contours, darkest indicating the weakest field strength).  The lines show the projection of the ForeCAT CME's toroidal axis with the color indicating the CME height.  This CME quickly deflects to the southeast, reaching a nearly constant position by 10 $R_{\odot}$, but shows little rotation.

We continue the ForeCAT simulations out to 50 $R_{\odot}$ to ensure we capture the full deflection and rotation.  We then use the position and orientation of the CME at 50 $R_{\odot}$ as inputs for the FIDO simulation.  Note that we do not perform any simulations between 50 $R_{\odot}$ and 1 AU, we assume there is no further deflection and rotation. Additionally, we make no attempt to simulate the arrival time of the ICME, it is assumed to start when the observed ICME starts.  Figure \ref{fig:ex}(h) compares the FIDO results for this case (dark blue) and the ACE and Wind in situ observations.  We also consider FIDO results driven by the COR2 GCS reconstruction (purple), as well as several best fit cases (green and light blue), discussed in Section \ref{BestFits}. 

\section{Coronal Results}\label{Cor}
We first analyze and compare the GCS reconstructions and ForeCAT simulations before investigating the in situ magnetic field near 1 AU as this set of 45 CMEs represents the largest study of both observations and simulations of CME deflection and rotations.

\subsection{Observations}\label{Obs}
\begin{figure}[h!]
\centering
\includegraphics[width=\textwidth]{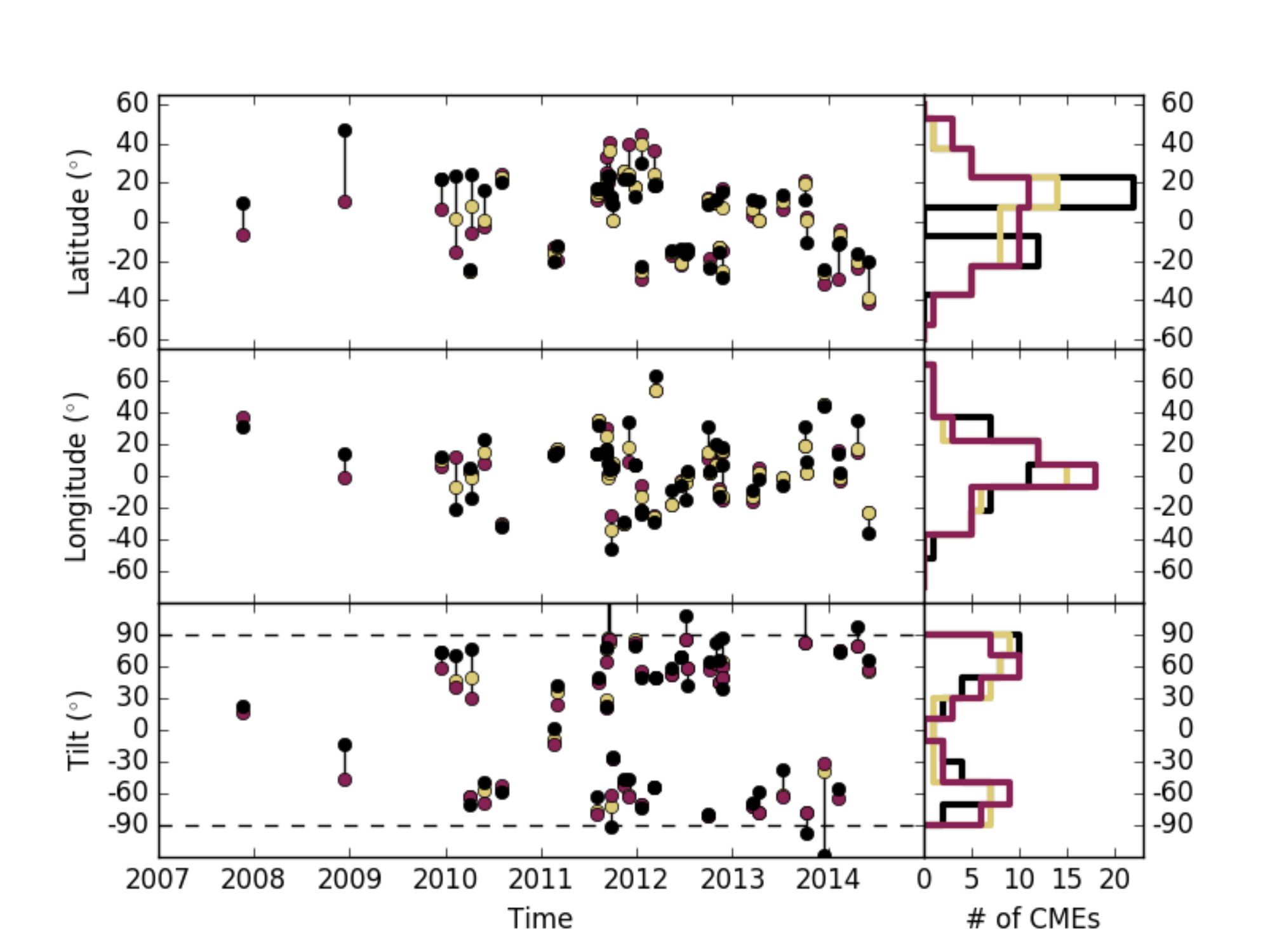}
\caption{Time series plot showing the evolution of each CME's latitude (top), Stonyhurst longitude (middle), and tilt (bottom).  The black circles indicate the values determined at the surface of the Sun, and the yellow and maroon circles correspond to the values from the COR1 and COR2 GCS reconstructions.  The black lines connect all the circle corresponding to a single CME.  The right panels show histograms of the surface, COR1, and COR2 values for the whole time series.}
\label{fig:histo}
\end{figure}

We first consider the deflections and rotations determined from the combination of the initial location and the coronagraph reconstructions.  Figure \ref{fig:histo} shows the latitude (top row), longitude (middle row), and tilt (bottom row) for each CME versus time.  Each circle represents a measurement at a different height with the black circles showing the location inferred from the PIL and the yellow and maroon circles showing the values from the COR1 and COR2 reconstructions.  The black lines connect the measurements for each CME.  The right panels contain histograms for the measurements at each height.

The top panels show that the majority of the CMEs initiate in two bands of latitude location at $\pm$10-20$^{\circ}$ latitude.  By the time that the CME reaches the COR1 and COR2 fields-of-view the distribution has flattened out, filling in the lowest latitudes and extending to higher latitudes.  Note that, despite this being a sample of Earth-impacting CMEs, we observe deflections both toward and away from the Earth's latitude.  Both directions occur with relatively equal frequency over the full time span.  Of the 45 CMEs, 20 deflect toward the Earth's latitude, and 25 deflect away with an average magnitude of 11.4$^{\circ}\pm$11$^{\circ}$ toward and 9.0$^{\circ}\pm$6$^{\circ}$ away, with the uncertainties representing the standard deviations.  The largest equatorward deflections occurred between 2008 and 2011, which corresponds to solar minimum.  This supports the trend of eruptive prominences being observed at higher latitudes than their associated CMEs seen during previous solar cycles (Figure 15.14 of \citet{Gop15EP}). During this time the CMEs initiate at higher latitudes, and the HCS is relatively flat and near the solar equator, leading to the large latitudinal deflections.  We futher discuss this and other solar cycle trends in Section \ref{sims}.

The middle panel shows the CME longitude relative to the Earth's location for each CME.  While some evidence may exist for a systematic eastward deflection of fast ICMEs \citep{Wan04} we do not expect the coronal forces to have a preferred direction of longitudinal deflection.  Since these are Earth-impacting CMEs, we may expect their longitude to move closer to Earth on average (therefore increasing the likelihood of impact).  The histograms in the middle panel do show a noticeable increase in CMEs in the closest longitude bins, 29 CMEs deflect toward the Earth's longitude.  We also, however, find some CMEs that move to very far longitudinal distances.  We find an average longitudinal deflection of 10.1$^{\circ}\pm$8$^{\circ}$ for the CMEs moving toward Earth, and 5.5$^{\circ}\pm$6$^{\circ}$ for the CMEs moving away.

In the bottom panel showing the orientation we have several CMEs that rotate across the boundary between $\pm$90$^{\circ}$.  For better visualization we show these cases as having an initial orientation with absolute magnitude greater than 90$^{\circ}$ so that the rotation appears as a continuous line.  For example, a CME that rotates clockwise from -80$^{\circ}$ to 80$^{\circ}$ would appear as a rotation from 100$^{\circ}$ to 80$^{\circ}$.  The histograms show the initial orientations without any adjustments.  The majority of the CMEs start with relatively vertical orientations, then tend to rotate toward more horizontal orientations.  We find an average rotation of 16.8$^{\circ}\pm$17$^{\circ}$ and both clockwise and counterclockwise rotations occur.  More CMEs rotate clockwise than counterclockwise (32 compared to 13) but the average clockwise rotation is comparable to the average counterclockwise rotation (17.7$^{\circ}\pm$14$^{\circ}$ versus 14.6$^{\circ}\pm$23$^{\circ}$)

Previous observations and simulations have shown that the largest deflections occur very close to the Sun \citep{Byr10,Isa14,Kay15}.  Between the initial location/orientation and the COR1 reconstructions we find average latitudinal deflections of 5.2$^{\circ}\pm$5$^{\circ}$, longitudinal deflections of 6.4$^{\circ}\pm$5$^{\circ}$, and rotations of 12.6$^{\circ}\pm$12$^{\circ}$.  In comparison, between the COR1 and COR2 reconstructions we find average latitudinal deflections of 4.3$^{\circ}\pm$4.8$^{\circ}$, longitudinal deflections of 2.2$^{\circ}\pm$4$^{\circ}$, and rotations of 3.2$^{\circ}\pm$5$^{\circ}$.  The majority of the longitudinal deflections and rotations happen before the distance of the COR1 reconstruction, but we see relatively equal average latitudinal deflection before and after this distance.  This is driven primarily by the large latitudinal deflections that occur near solar minimum, CMEs occuring in 2010 or earlier have an average latitudinal deflection of 8.5$^{\circ}\pm$7$^{\circ}$ between the COR1 and COR2 reconstructions, compared to only 3.5$^{\circ}\pm$4$^{\circ}$ for CMEs from 2011 or later.  Between the COR1 and COR2 reconstructions, the average latitudinal deflection for CMEs occuring after 2010, as well as the average longitudinal deflection and rotation for all cases, is comparable to the uncertainties in the reconstruction technique.

\subsection{Simulations}\label{sims}
\begin{figure}[h!]
\centering
\includegraphics[width=\textwidth]{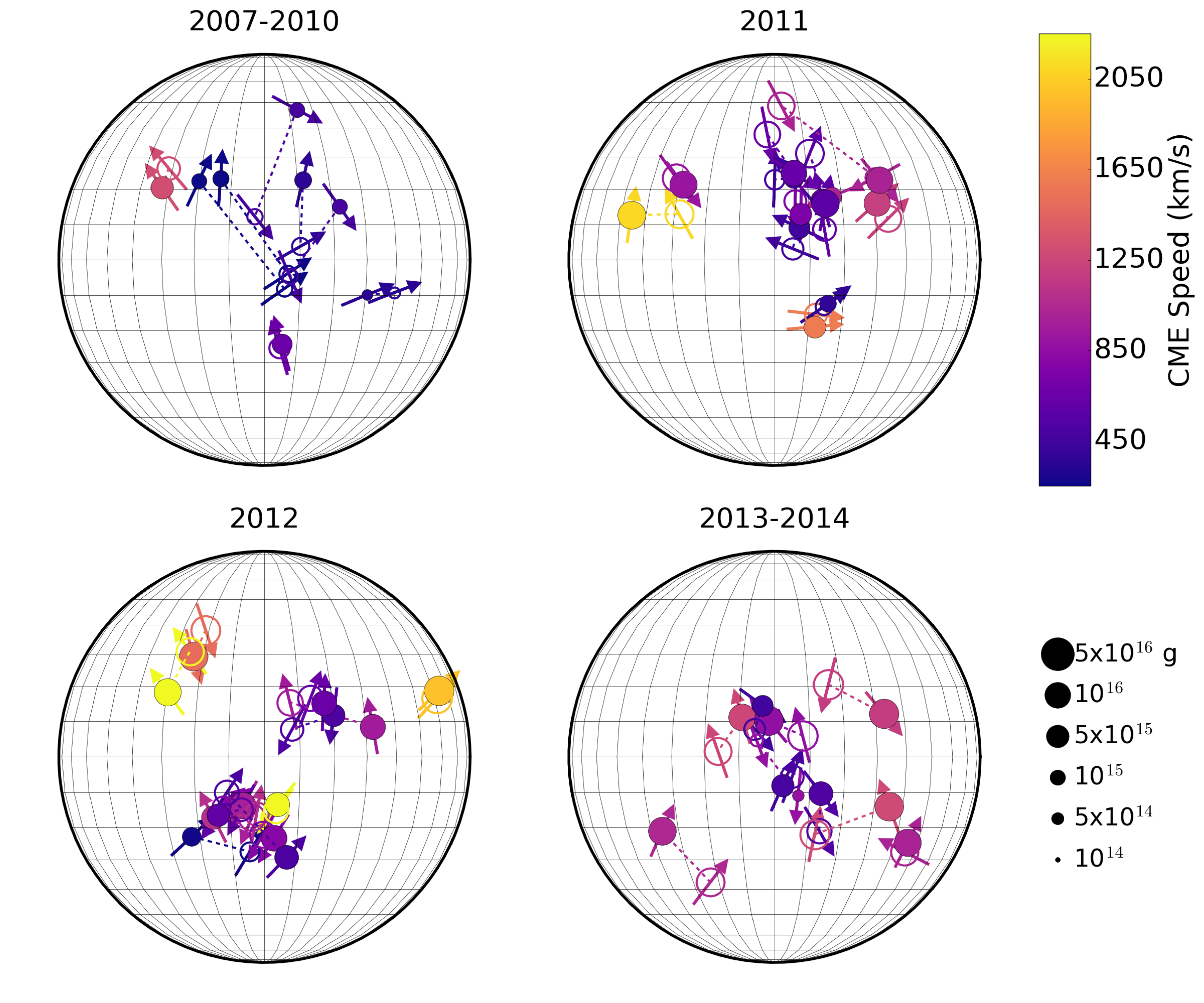}
\caption{Initial and final position and orientation of each CME.  The circle location shows a CME latitude and longitude projected onto the solar disk and the bar indicates the orientation of a CME's toroidal axis, with the arrow indcating the direction of the toroidal magnetic field.  Filled circle indicate the initial values used in ForeCAT and empty circles represent the ForeCAT results at 50 $R_{\odot}$ with a dashed line connecting each pair.  The color and symbol size indicate the coronal velocity and mass according to the scales on the right. The 45 CMEs have been split by eruption time into four roughly even groups simply for clarity.}
\label{fig:circs}
\end{figure}

We next look at the ForeCAT results.  Figure \ref{fig:circs} shows the initial (filled circles) and final position (empty circles, at 50 $R_{\odot}$) projected onto the solar disk.  The bar show the orientation of the CME at each distance with the arrow indcating the direction of its toroidal magnetic field.  The size of each circle indicates the CME mass and the color represents the maximum radial speed in the simulation. The simulated CMEs are split between four panels according to time for clarity.  The thin black lines in the background mark every 10$^{\circ}$ in latitude and longitude.

We note that Figures \ref{fig:histo} and \ref{fig:circs} respectively show the observed and simulated CME behavior.  Nearly all the ForeCAT results match the GCS reconstructions within their uncertainties, but the final latitudes and longitudes vary by 3$^{\circ}$ and the final tilts vary by 6$^{\circ}$, on average.  This difference between the two sets is sufficiently small that both figures would not visually change significantly were the other data set used.  While the differences may be small on visual scales, we later further discuss the exact magnitude of these differences and show that these small differences can yield significantly difference in situ profiles.

The top left panel shows the simulations corresponding to CMEs during solar minimum.  Comparison with the other panels shows that these CMEs tend to have the smallest masses and radial velocities.  As seen in Figure \ref{fig:histo}, these solar minimum CMEs start at higher latitudes and exhibit large latitudinal deflections toward the equator.  The magnetic background is weakest during solar minimum, but the low CME masses and speeds allow the weaker forces to still yield large deflections.  For these CMEs we find an average latitudinal deflection of 17.4$^{\circ}\pm$12$^{\circ}$, longitudinal deflection of 11.3$^{\circ}\pm$9$^{\circ}$, and rotation of 21.4$^{\circ}\pm$19$^{\circ}$.

As the solar cycle progress (top right then bottom left and bottom right panels), the CME masses and speeds increase, and the latitudinal deflections decrease.  In each panel the average masses are 1.8$\pm$1.4$\times$10$^{15}$ g, 7.8$\pm$5$\times$10$^{15}$ g, 9.4$\pm$6$\times$10$^{15}$ g, and 12$\pm$8$\times$10$^{15}$ g and the average speeds are 515$\pm$300 km s$^{-1}$, 896$\pm$480 km s$^{-1}$, 1106$\pm$690 km s$^{-1}$, and 896$\pm$300 km s$^{-1}$.  The average magnetic field strength of the background also increases during this time, and the balance between the increases in the masses, speeds, and forces determines the change in the magnitude of the deflection and the rotation. For the 2011, 2012, and beyond 2013 groups we find average latitudinal deflections of 6.8$^{\circ}\pm$7$^{\circ}$, 4.9$^{\circ}\pm$5$^{\circ}$, and 8.5$^{\circ}\pm$5$^{\circ}$, average longitudinal deflections of 7.0$^{\circ}\pm$9$^{\circ}$, 8.1$^{\circ}\pm$7$^{\circ}$, and 8.5$^{\circ}\pm$7$^{\circ}$, and average rotations of 14.0$^{\circ}\pm$15$^{\circ}$, 9.3$^{\circ}\pm$10$^{\circ}$, and 26.8$^{\circ}\pm$26$^{\circ}$.  For the 45 CMEs in this paper, we find that the largest deflections occur during solar minimum due to the low CME masses and velocities.  This is followed by a brief decrease in the deflection as the CME masses and velocities tend to decrease, until the background magnetic forces increase sufficiently to balance the more massive, faster CMEs.  We emphasize that this corresponds to the average behavior over our sample, and that individual CMEs' behavior may differ.

\subsection{Trends in Deflection and Rotation}
The average simulated latitudinal and longitudinal deflection are comparable over the full sample (8.5$^{\circ}$ and 8.4$^{\circ}$, respectively; 9.1$^{\circ}$ and 8.3$^{\circ}$ in observations, with higher uncertainty).  The ratio of the latitudinal and longitudinal deflection varies over time because the average longitudinal deflection remains relatively constant, but the latitudinal deflection is largest near solar minimum.  Similar to the deflection, we also find that the largest rotations also occur at the beginning and end of our sample.  For this set of CMEs we find that during solar minimum and maximum the balance of CME mass/speed and the magnetic forces/torques favors larger deflections and rotations but during the rise phase the masses/speeds have increased more proportional to the forces/torques leading to smaller deflections and rotations.

As mentioned in section \ref{Obs}, and also seen in the ForeCAT simulations, not all CMEs deflect toward the Earth's location, despite all corresponding to Earth-impacting ICMEs.  \citet{Xie13} show a similar figure to Figure \ref{fig:circs}, which shows that MCs tend to deflect toward the Earth whereas non-MCs are more likely to deflect away.  From the simulations, 24 CMEs deflect away from the Earth's latitude, and 16 CMEs deflect away from the Earth's longitude.  Note that we find one CME where the direction of the latitudinal deflection varies between the observations and simulations and three where the direction of the longitudinal deflection varies; all these cases have small (<5$^{\circ}$) deflections.  Of the CMEs deflecting away from the Earth, roughly half have a deflection away greater than 5$^{\circ}$ (13 in latitude and 5 in longitude).  Additionaly, only two CMEs have both the latitude and longitude simultaneously move away from that of the Earth, and for only one of these does the deflection exceeds 5$^{\circ}$ in both directions (CME 13).

The CMEs exhibit both westward and eastward deflections and we see no hemispheric trends in the direction of the longitudinal direction (17 out of 27 deflect eastward in the north, 9 out of 18 in the south, from the simulations).  Similarly, no trends are seen in the direction of the latitudinal deflection (14 out of 27 deflect equatorward in the north, 8 out of 18 equatorward in the south).  

Observations and simulations suggest that CMEs with positive/negative helicity rotate clockwise/counterclockwise \citep[e.g.][]{Gre07, Lyn09}.  This rotation can be explained via conservation of helicity and an ``unraveling'' of the flux rope as it erupts.  For CMEs that follow the \citet{Bot94} polarity (negative/positive helicity in the north/south) we would expect counterclockwise rotation in the north and clockwise rotations in the south.  However, as mentioned in section \ref{ObsMeth}, the \citet{Bot94} relation is only valid for average CME properties and not necessarily specific CMEs due to the complicated structure of the ARs.  We find that about 60\% of CMEs rotate as expected for their hemisphere - 17 of the 27 CMEs rotate counterclockwise in the north and 11 of the 18 CMEs rotate clockwise in the south.  Considering the polarity of the CMEs used for the FIDO simulations, 6 of the 19 left-handed CMEs rotate counterclockwise and 11 of the 26 right-handed CMEs rotate clockwise.  ForeCAT only includes rotation from the background forces, not from the forces within the CME itself.  These external forces yielding very different results than what is expected from the internal forces, but we note that the ForeCAT simulations are consistent with the GCS reconstructions.  Only 17 of the 45 CMEs have a rotation exceeding the 10$^{\circ}$ uncertainty in the reconstructed orientation, illustrating the difficulty of accurately determining CME rotations from coronal reconstructions.

\begin{figure}[h!]
\centering
\includegraphics[width=\textwidth]{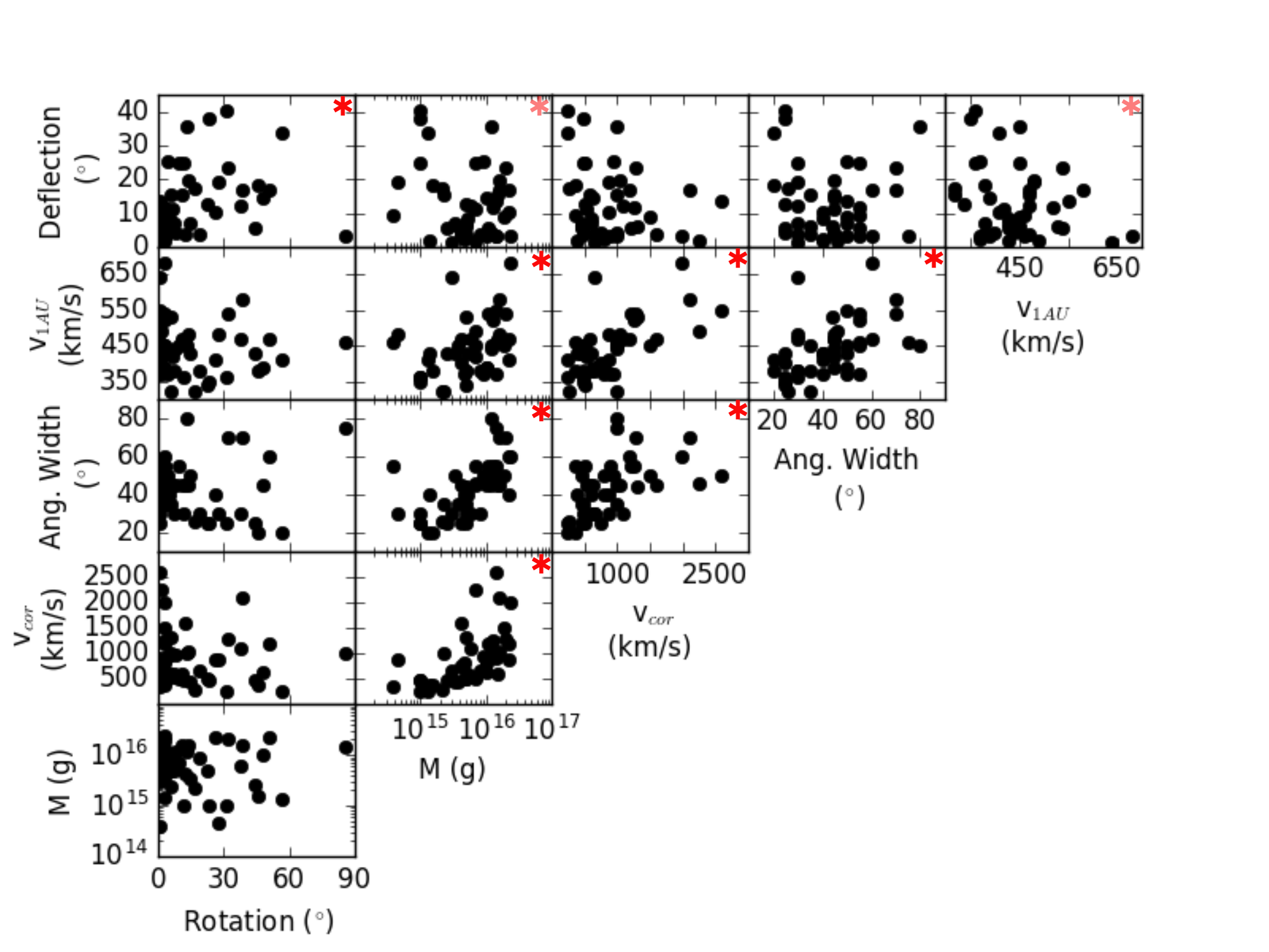}
\caption{Correlation plot for parameters related to large-scale CME properties (speed, mass, and angular width) and the total deflection and rotation.}
\label{fig:defcor}
\end{figure}

In addition to looking for solar cycle or hemispheric trends, we can also compare the deflection and rotation (from the ForeCAT simulations) with various CME properties.  Figure \ref{fig:defcor} shows correlations between the deflection, rotation, mass, $M$, coronal speed, $v_{cor}$, angular width, and speed at 1 AU, $v_{1AU}$, for all 45 CMEs.  Note that the correlations are determined with respect to the base-10 log of the CME mass due to the values spanning several orders of magnitude.  

\begin{table}
\caption{Correlations for the parameters shown in Figure \ref{fig:defcor}.}	
\begin{center}
\begin{tabular}{ c|c } 
	Parameters & r \\
 \hline 
 Def, Rot &  0.300\\ 
 Def, $M$ &  -0.246\\ 
 Def, $v_{cor}$ &  -0.212\\ 
 Def, AW & -0.059\\
 Def, $v_{1AU}$ & -0.238 \\
 $v_{1AU}$, Rot & -0.068\\
 $v_{1AU}$, $M$ & 0.365 \\
 $v_{1AU}$, $v_{cor}$ & 0.624 \\
 $v_{1AU}$, AW & 0.482\\
 AW, $M$ & 0.618\\
 AW, $v_{cor}$ & 0.518\\
 $v_{cor}$, $M$ & -0.119\\
 $M$, Rot & 0.003 \\
 \hline
\end{tabular}
\label{tab:defcor}
\end{center}
\end{table}

Our results reproduce many of the trends from large catalogs such as LASCO, for example the positive correlation between mass, speed, and size \citep[e.g.][]{Yas04, Vrs07b}.  For our set of 45 CMEs, a correlation is significant at a p-value of 0.05 for a correlation coefficient of 0.248 or greater.  Table \ref{tab:defcor} shows the correlation values, $r$, for the pairs of parameters shown in Figure \ref{fig:defcor}.  In Figure \ref{fig:defcor} the panels with a statistically significant correlation are marked with a red "*" in the top right corner, with the pale red markers indicating borderline correlations.  For the CME properties we find significant correlations between the coronal and 1 AU speeds, the mass and angular width, the mass and the coronal speed, the coronal speed and the angular width, the 1 AU speed and the angular width, and between the mass and 1 AU speed.  In a study of the geoeffectiveness of Solar Cycle 23 CMEs, \citet{Gop08Geo} also found strong correlations between the coronal and 1 AU speeds, and these speeds and the magnetic field at 1 AU.

All of the correlations with the CME deflection and rotation are less significant than the correlations amongst the CME properties themselves.   We find that the deflection and rotation are significantly correlated (0.300), and that the mass and total deflection are inversely correlated at just about the p$=$0.05 significant level.  We find inverse correlations just below the p$=$0.05 significant level for the the coronal and 1 AU speeds and the deflection.   All other correlations are much below the significant level.  These weaker correlations are likely due to number of factors that influence the deflection and rotation.  Specifically, as the solar cycle progresses from solar maximum, the solar magnetic field strength, and therefore the deflection forces, increases, but simultaneously so do the average CME mass, speed, and angular width, counteracting the increase in the forces. 

\section{Near-Earth Results}\label{1AU}
We now look at the comparison between the FIDO results and the in situ observations.  First, in section \ref{score}, we define a score that we will use to determine the quality of the model fits to the in situ observations.  In section \ref{FIDO1} we see how well FIDO, driven by ForeCAT results, reproduces the in situ observations and compare with FIDO results driven by the GCS reconstructions instead of the ForeCAT results.  In Section \ref{BestFits}, we look for the best fits possible with the simple flux rope model to better understand the difference between how well we can determine the model inputs using either ForeCAT results or GCS reconstructions and the limitations of the simple model itself.

\subsection{Goodness-of-Fit Score}\label{score}
FIDO uses a simple flux rope model, which will never be able to reproduce the small scale structure of flux ropes that results from dynamic effects such as turbulence or reconnection.  Our focus is to predict the correct polarity and relative magnitude of the three components ($B_x$, $B_y$, and $B_z$) on order of hourly time scales.  Accordingly, we calculate the goodness-of-fit score using the hourly averages of both the FIDO results and the ACE in situ data.  The scores do not vary significantly whether ACE or Wind data is used.  

We first determine the difference between the hourly averages of the simulations and observations and for each component
\begin{linenomath*}
\begin{equation}\label{eq:Deltai}
\delta_i = |B_{FIDO,i} - B_{obs,i}|
\end{equation}
\end{linenomath*}

where $i=x$,$y$,$z$.  We determine the full vector magnitude of the hourly error
\begin{linenomath*}
\begin{equation}\label{eq:Delta_unnorm}
\delta = \sqrt{\delta_x^2 + \delta_y^2 + \delta_z^2}
\end{equation}
\end{linenomath*}

The importance of the magnitude of $\delta_i$ depends on the magnitude of the actual flux rope magnetic field - a larger $\delta_i$ is more acceptable for a flux rope with a stronger magnetic field, however we want to be able to compare between different ICMEs.  To normalize $\delta_i$, we use the average of the observed hourly magnetic field strength, which is equivalent to setting $B_{FIDO,i}=0$ in Equation \ref{eq:Deltai}.  We normalize both the scores for the individual components and the total error by the average observed magnetic field.  If the components are individually normalized by their own average field strength then extremely high score can occur when an individual component has a small magnitude, making it difficult to compare between the three vector components.

We calculate the average values of $\delta_i$ and $\delta$ to determine how successfully the model matches the individual components and the CME as a whole.  For the total score, a score of zero corresponds to a perfect fit and a score of one corresponds exactly to assuming either no magnetic field or twice the magnitude for the flux rope duration.  In this work we take a score of one as a lenient upper limit for a ``good'' fit, keeping in mind that we scale the simulations such that the average simulated total magnitude match the observed total magnitudes.

Additionally, we penalize results which yield ICME durations that are either too short or too long.  The simulated front end is forced to match the observed flux rope front, and we tend to use the most conservative (i.e. earliest) estimate of the end.   Accordingly, we reject FIDO results with a back end more 30 minutes earlier than the observed value.  We are less strict with cases that extend beyond the observed back end as it is often difficult to define observationally, and penalize by adding to the score 0.1 times the fractional hours the model extends past an hour beyond the observed end point.

\subsection{FIDO-driven and GCS-driven results}\label{FIDO1}
\begin{figure}[h!]
\centering
\includegraphics[width=\textwidth]{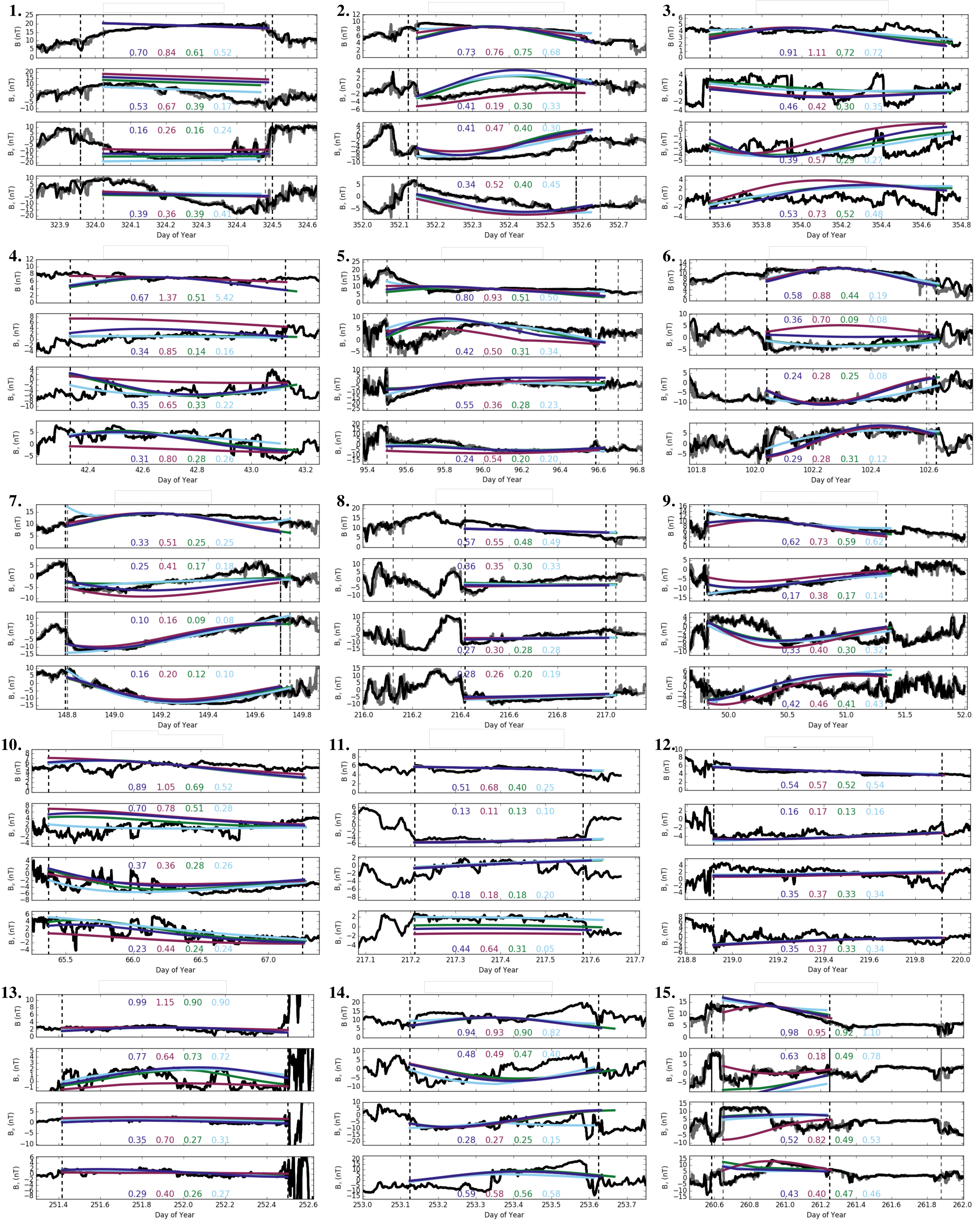}
\caption{Comparison between the ACE and Wind in situ observations (black and gray lines) and ForeCAT-driven FIDO results (dark blue), the best fit force free and circular FIDO results (green and light blue, respectively), and GCS-driven FIDO results (purple) for the first 15 CMEs.  The black and gray dashed lines respectively indicate the CME start and stop times according to the Richardson and Cane and WIND ICME lists.  For each CME, the top panel shows the total magnetic field, $B$, followed by $B_x$, $B_y$, and $B_z$.  The number in the $B$ panel indicate the score for each of the four FIDO models and the other three panels show the individual score for each of the vector components.}
\label{fig:multi1}
\end{figure}

\begin{figure}[h!]
\centering
\includegraphics[width=\textwidth]{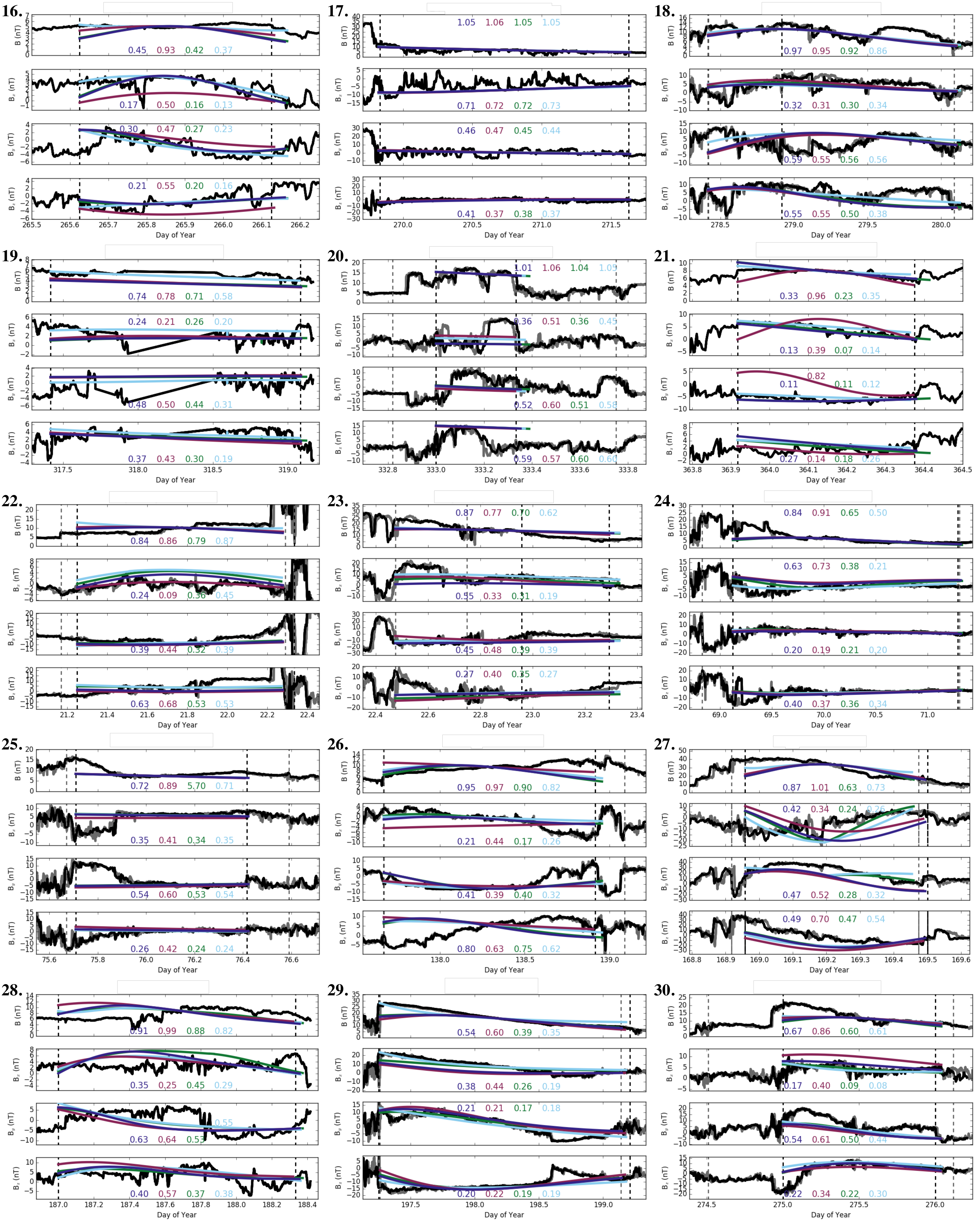}
\caption{Same as Figure \ref{fig:multi1} but for the middle 15 CMEs.}
\label{fig:multi2}
\end{figure}

\begin{figure}[h!]
\centering
\includegraphics[width=\textwidth]{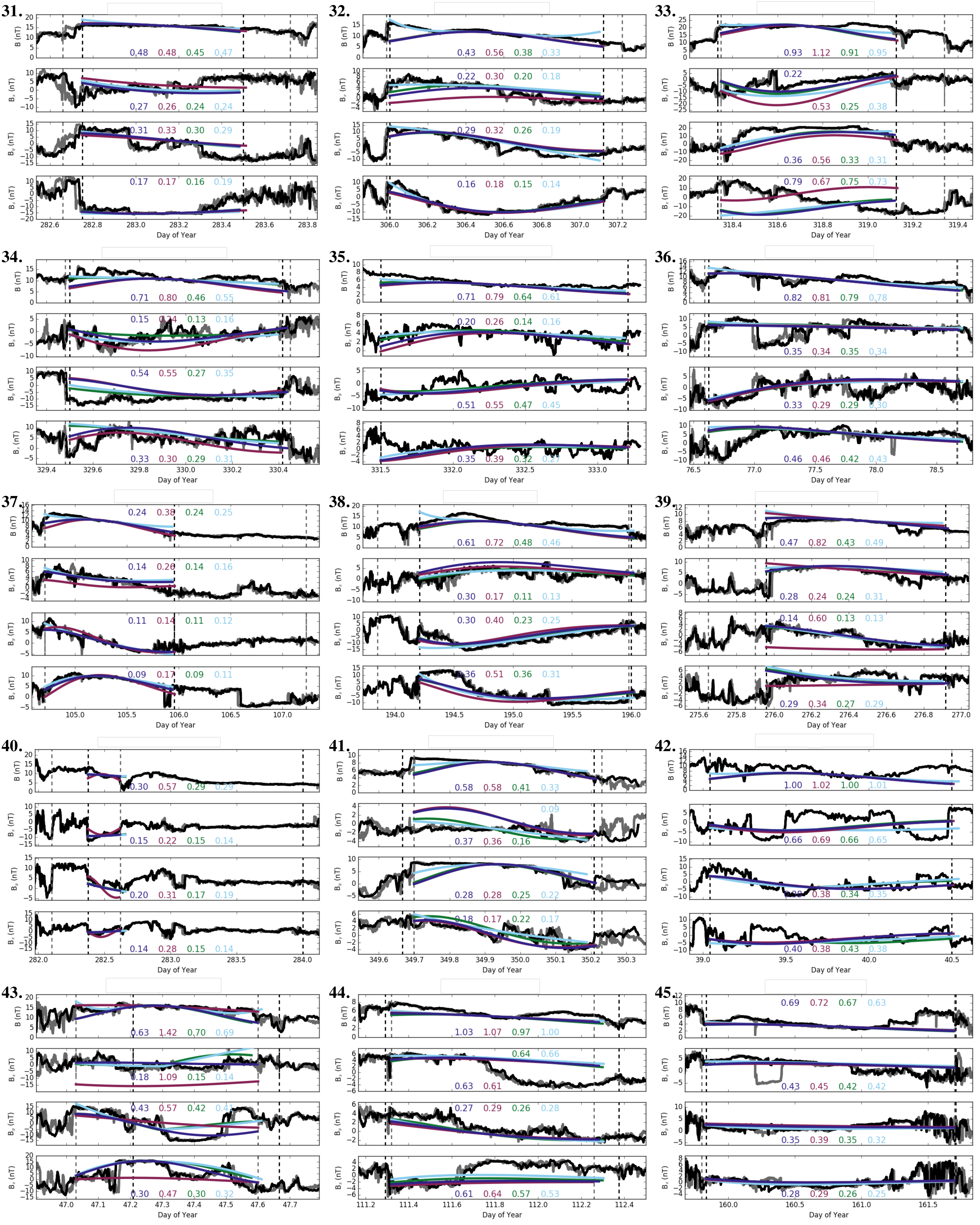}
\caption{Same as Figure \ref{fig:multi1} but for the last 15 CMEs.}
\label{fig:multi3}
\end{figure}

Figures \ref{fig:multi1}-\ref{fig:multi3} compare the FIDO results with observations for each of the 45 CMEs (15 flux ropes in each figure, grouped according to time).  The ID of the eruption is shown at the top of every panel.  Each panel shows, from top to bottom, the total magnitude $B$, and $B_x$, $B_y$, and $B_z$ in GSE coordinates.  The black and gray lines correspond to the in situ data from ACE and Wind, respectively, and the black and gray dashed vertical lines indicate the ICME start and stop times from RC and Wind ICME lists.  Here we focus on fitting the flux rope or flux-rope-like portion of the ICME not the shock or sheath.  The dark blue line shows the ForeCAT-driven FIDO results and the purple line shows the GCS-driven results.  The green and light blue lines are best fit results, discussed in section \ref{BestFits}.  Finally, the numbers at the bottom of the $B$ panel show the full score for each of the four models, and the score for each of the individual components are shown in the $B_x$, $B_y$, and $B_z$ panels.

We find that FIDO does a reasonable job of reproducing the in situ magnetic field with the average total score for all 45 CMEs being 0.72$\pm$0.22, with the uncertainty corresponding to the standard deviation.  For all but 3 CMEs (CMEs 17, 20, and 42) we obtain acceptable fits, defined by a score less than one.  If we separate the ``good'' and ``bad'' fits, dividing at a score of one, we get averages of 0.79$\pm$0.20 and 1.03$\pm$0.02 for the two groups.  The average good score changes very little from the full average due to the low number of bad fits, and we find that the bad fits are not significantly greater than one and therefore wildly inaccurate, rather just slightly worse than assuming zero magnetic field or twice that of the actual CME.  For the full sample the score varies between 0.24 and 1.05.

We can also look at the scores for the individual vector components.  Accurately reproducing the $B_z$ component is the most essential for accurate space weather predictions (technically in Geocentric Solar Magnetospheric coordinates, though the difference from GSE is small). FIDO, however, puts no special emphasis on this component - the vector compenents simply result from the CME orientation and the distance from the toroidal axis.  This is reflected in the scores for the individual components.  All three components have an average value of 0.35 with standard deviations varying between 0.14 and 0.18, suggesting that FIDO overall does equally well for all components.

Despite using a simple flux rope model, we see quite a bit of variance in the resulting in situ profiles.  Many of the cases show a smooth rotation in the magnetic field in one or more parameters, whereas others exhibit less variation and have all the components maintain a nearly constant value over the duration.  The profiles depend most strongly on the distance from the synthetic spacecraft to the flux rope axis with the flattest profiles corresponding to the farthest distances or flank hits.

For several cases we find that FIDO yields an acceptable fit to two of the magnetic field components, but the score is driven to a higher value by an extremely poor fit in the other parameter.  The most extreme example of this is the 09 November 2012 CME (CME 33) where FIDO reproduces $B_x$ and $B_y$ relatively accurately but completely fails to produce the behavior of the $B_z$ component.  More commonly, the score is driven to higher values by mediocre fits to two or three of the components rather than an extreme in one component (e.g. CME 14 on 06 September 2011 and  CME 31 on 05 October 2012).  In some cases, we see more of an east-west rotation in the FIDO results compared to a more significant north-south rotation in the observations (e.g. CME 18 on 10 October 2011).  This is most likely results from impreciseness in the orientation of the CME, which can rotate the magnetic field between the nonradial directions. For the best fits FIDO simulataneously reproduces all components (e.g. CME 16 on 19 September 2011 and CME 29 on 12 July 2012).  For the worst fits FIDO fails at reproducing all components for a significant fraction of the flux rope duration (e.g. CME 15 on 14 September 2011 and CME 20 on 26 November 2011) suggesting that the simple flux rope model is not sufficient or appropriate for these cases.

Since we are using a simple flux rope model that does not include any effects of the deformation or deterioration that may occur between the Sun and 1 AU, we expect that FIDO will do best on ICMEs that appear the most flux-rope like.  We visually assign each ICME a quality score based on how flux-rope like they appear.  A quality of 1 indicates an obvious flux rope with a smooth rotation in the magnetic field (e.g. CME 7 on 24 May 2010 or CME 32 on 27 October 2012) and a quality of 3 indicates little-to-no obvious structure that can be discerned from the background magnetic field (e.g. CME 4 on 07 February 2010 or CME 17 on 24 September 2011), and quality 2 falls somewhere in between.  The qualities are listed in Table 2 of the supplementary material.  We emphasize that these qualities are meant simply to allow for a better understanding of when FIDO works best and not meant to be used as a precise grouping of ICMEs based on rigorous criteria for other studies.  

From the 45 CMEs we define 8 CMEs as quality 1, 15 CMEs as quality 3, and the remaining 22 as quality 2.  Of the three bad fits, two of the CMEs fall in group 3 and one in group 2.  We find an average score of 0.55 for the quality 1 ICME's, 0.74 for the quality 2 ICMEs, and 0.77 for the quality 3 CMEs.  FIDO clearly does the best for the quality 1 ICMEs, but all qualities have a standard deviation of 0.2 in their average score showing there is a wide dispersion in the scores in any single group.  

As mentioned in section \ref{models}, rather than keeping $B_0$ as a free parameter, we automatically normalize the FIDO magnetic field using the middle four hours of the ICME passage.  We can compare with the $B_0$ that produces the best score for each case.  For some of the ICMEs, which tend to have the poorest fits, the best score comes from a $B_0$ that does not visually match the observed magnitude.  We exclude these cases and find an average score of 0.63 for the remaining 31 ICMEs using the best fit $B_0$.  These same ICMEs have an average score of 0.64 using the automatically normalized $B_0$, suggesting that using this method to remove $B_0$ as a free parameter is reasonable.

We also consider the FIDO results driven by the latitude, longitude, and tilt from the COR2 GCS reconstruction instead of the ForeCAT values.  These results are shown in maroon in Figures \ref{fig:multi1}-\ref{fig:multi3}.  We use the same angular width as the ForeCAT-driven FIDO results, but allow for changes in the shape ratios $A$ and $B$ as they are not constrained by any of the observations.

On average, the GCS-driven results tend to do slightly worse at reproducing the observed in situ magnetic field.  The GCS-driven results have an average score of 0.86$\pm$0.22 (compared to 0.72 for all the ForeCAT-driven cases).  Whereas the ForeCAT-driven cases resulted in scores greater than one for 3 of the 45 ICMEs, we find that 11 of the GCS-driven ICMEs yield score greater than one.  If we compare the good and bad scores, the GCS-driven ICMEs have an average good score of 0.77$\pm$0.20 compared to the 0.69$\pm$0.20 for average good ForeCAT-driven cases and the average bad GCS-driven score is 1.13$\pm$0.13 compared to 1.03$\pm$0.02 fo the ForeCAT-driven results.  Both sets have significantly higher average GCS-driven scores than the average bad ForeCAT-driven score.

The difference between the average good and bad scores for the GCS- and ForeCAT-driven results highlights the inheritent difficulty in using the GCS reconstructions.  These reconstructions do not typically have a systematic error or bias in their values, rather just a large uncertainty due to the difficulty of making precise visual fits to white-light images and the degeneracy of parameters that can yield visually-acceptable fits.  The majority of the GCS-driven cases have scores only slightly worse than the ForeCAT-driven results.  Sixteen of the cases have GCS-driven scores within 0.05 of the ForeCAT-driven score, and 1 case (CME 23) actually has a better score than the ForeCAT-driven results.  However, a much larger, significant portion of the GCS-driven results yield bad fits.  The GCS-results can certainly yield good fits to observations, but it has a much higher risk of returning wildly inaccurate results. 

\subsection{Best Fits}\label{BestFits}
Section \ref{FIDO1} shows that FIDO is capable of reproducing the in situ magnetic field for the majority of the considered ICMEs and that it is preferrable to use ForeCAT results rather than GCS reconstructions to determine the FIDO input parameters, but it is unclear whether a different set of parameters could yield a better fit.  To determine this we look for the best fit using FIDO's simple flux rope and torus shape.  We want to determine the limitations of how well the model itself can reproduce the in situ results so that we can better understand our ability to determine its input parameters with ForeCAT.

We perform a random walk, keeping or rejecting new parameters based on whether they improve the score.  We find that 4000 steps tends to be long enough to converge upon a best fit.  To narrow down the region of parameter space explored we restrict the parameters to within 5$^{\circ}$ in latitude and longitude, 10$^{\circ}$ in tilt, and 10$^{\circ}$ in angular width.  We force $A$ to remain in between 0.4 and 1 and $B$ to be less than $A$. 

We use both the force free flux rope model, used for all previous results in this paper and \citet{Kay17FIDO} and the circular flux rope model \citep{Nie16}.  We continue to use the automatic normalization of the total magnetic field strength so there are no additional parameters in the random walk for the force free model, but we allow $C_1$ and $\tau$ of the circular model to vary between 0.5 and 2.  Figures \ref{fig:multi1}-\ref{fig:multi3} show the force free and circular best fits in green and light blue, respectively. The best fit parameters are shown in Table 3 of the supplementary material.

The force free best fit scores consistently show an improvement over the ForeCAT-driven results.  The force free best fits have an average score of 0.63$\pm$0.22, but we still have two ICMEs with a score greater than one (CME 42 improves to good).  We begin each of the circular flux ropes with a $C_1$ of 1 and a $\tau$ of 1.5, which are average values for these parameters (Nieves-Chinchilla, private communication).  The average of the circular best fits is 0.61$\pm$0.24, comparable to the force free best fits, and four ICMEs still have a score greater than one (CME 15 is the additional bad fit with the circular model).  

We see a slight variation in the improvement of the individual vector components, though the standard deviations are much larger than the variation between components.  The $B_x$ score decreases to 0.29$\pm$0.17 for both the force free and circular models, $B_y$ to 0.31$\pm$0.12 and 0.30$\pm$13, and $B_z$ to 0.34$\pm15$ and 0.32$\pm$0.15 (previously about 0.35 for all components).  When the impact occurs near the CME nose the x-component is small as both the toroidal and poloidal field is in the yz-plane.  For these cases the score can often be easily improved by small changes in the position of the CME, bringing the impact point closer to the nose and minimizing the error, which may explain the slightly larger improvement in the x component.

In general, it appears that the circular model reproduces the the in situ magnetic field slightly better than the force free flux rope, however it does have additional free parameters.  Directly comparing the scores from the two best fits we find that 25 of the cases have scores within 0.05 of one another.  Of the remaining 20, the circular model does better for 14 cases, and the force free does better for 6 cases.

The circular model does tend to provide a better visual fit to the total magnetic field strength, particularly near the front and back of the flux rope because the parameter $C_1$ can adjust the ratio of the poloidal and toroidal field.  The match to the total magnetic field strength, however, does not explicitly factor into the score and we find that this does not necessarily translate into a better match to the indidual vector components.  

This is well-illustrated by the 15 Feb 2011 ICME (CME 9).  The circular flux rope better reproduces the enhanced magnetic field at the front of the ICME.  This leads to a better fit in $B_x$ (0.14 for circular versus 0.17 for force free) but the $B_y$ and $B_z$ scores are slightly worse, leading to an overall worse score for the circular model than the force free model (0.62 versus 0.59) despite the better apparent fit to the total magnetic field strength.

\begin{figure}[h!]
\centering
\includegraphics[width=\textwidth]{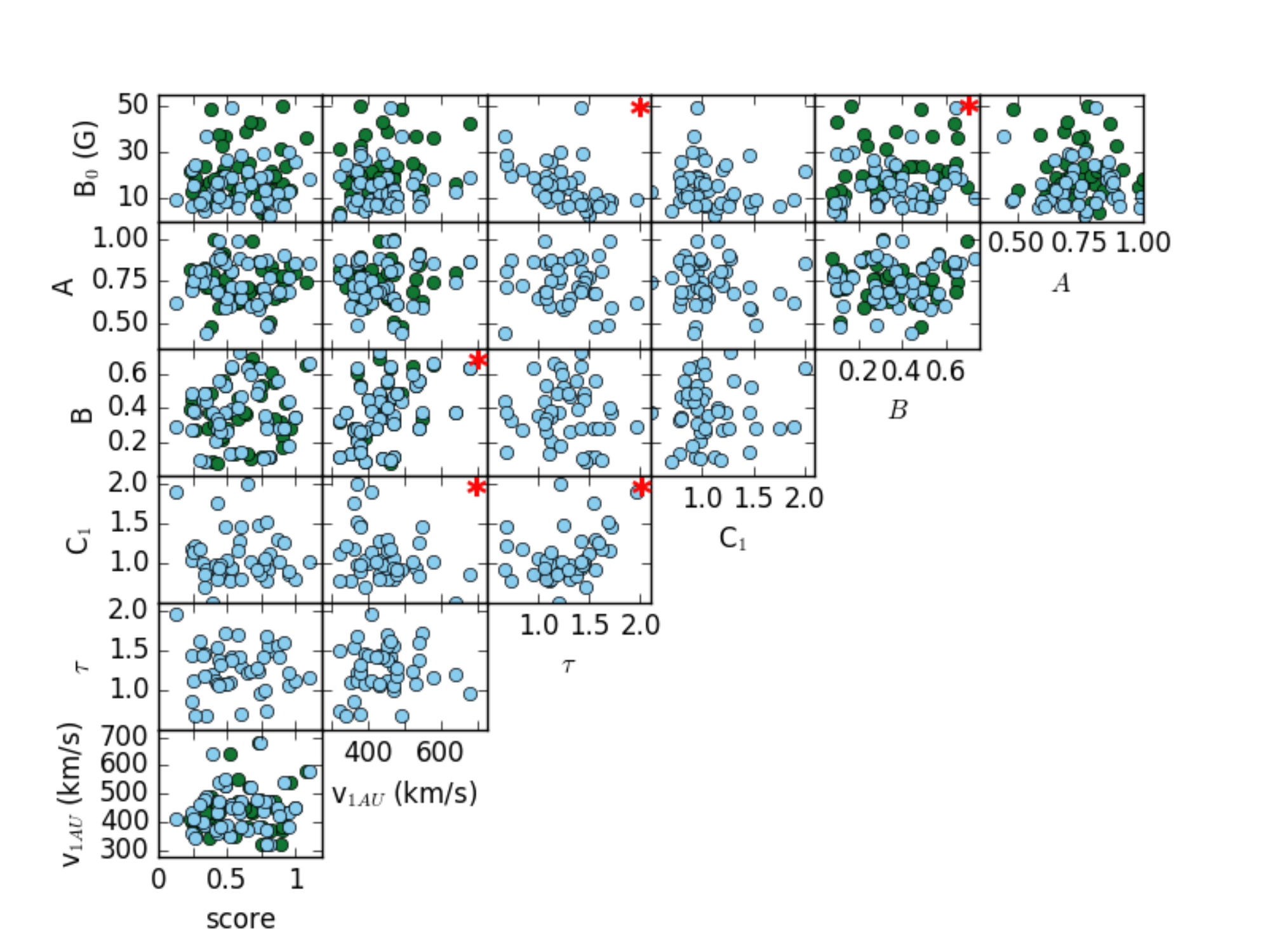}
\caption{Correlation plot for the parameters related to the FIDO flux rope.  The green circles show results from the force free best fits and the light values show results from the circular best fits.}
\label{fig:paramcorr}
\end{figure}

Similar to looking for correlations in the CME parameters in Figure \ref{fig:defcor}, we can also look for correlations between the FIDO input parameters.  Whereas we were able to reproduce many of the known correlations between CME properties in Figure \ref{fig:defcor}, here we seek to find new relationships between parameters that are often difficult to determine from remote observations.  Any significant correlations could help reduce the number of free parameter and facilitate future predictions of the in situ magnetic field.  Figure \ref{fig:paramcorr} shows the correlations between the total magnetic field strength $B_0$, the shape parameters $A$ and $B$, the circular model parameters $C_1$ and $\tau$, the velocity at 1 AU, $v_{1AU}$, and the fit score.

The y-axis of the top row shows the total magnetic field strength.  For these comparisons we use the value of $B_0$ that corresponds to the automatically-scaled value.  Both $\tau$ and $C_1$ affect the magnitude of the circular flux rope, so we typically find different values of $B_0$ for the force free and circular best fits.  On average, the force free $B_0$ tends to be about 1.59 times the circular $B_0$.  We find that $B_0$ and $\tau$ show a strong negative correlation (-0.486) for the circular model.  Both flux rope models also show correlation between $B_0$ and the cross-sectional shape parameter $B$ (0.305 and 0.311 for force free and circular, respectively).  None of the other FIDO parameters show significant correlations with $B_0$, nor do we find any correlation with the CME mass, angular width, or coronal velocity.

For the remaining combinations of parameters we see very few significant correlations.  The speed at 1 AU and $B$ are strongly correlated for both flux rope models (0.460, 0.478 for force free and circular, respectively).  We also find that the circular flux rope parameter $C_1$ is negatively correlated with the shape parameter $A$ (-0.312) and positively with $\tau$ (0.574).  All other correlations are below the 0.05 significant level.

We find no correlations between any of the FIDO model parameters and the quality-of-fit score, suggesting that there is not a systematic bias toward FIDO doing better for a certain type of ICME (e.g. faster, larger, stronger magnetic field).  We can also compare the score with the distance between the synthetic spacecraft/Earth and the ICME nose to see if there is a preference for nose or flank hits.  Using any of the FIDO scores (ForeCAT-driven or the force free or circular best fits) we find a positive correlation just at the 0.05 statistically significant level, suggesting that FIDO may do better for encounters near the nose as opposed to near the flanks.

For some of the parameters it is worthwhile considering their average values from our sample.  The precise values of the circular flux rope parameters have not yet been well-established, $C_1=1$ and $\tau=1.5$ are suggested initial values.  We confirm that these suggestions tend to be the appropriate values for these CMEs finding average values of 1.06$\pm$0.25 and 1.35$\pm$0.26 for $C_1$ and $\tau$, respectively.  \citet{Jan13,Jan15} have statistically analyzed ICMEs via their in situ magnetic field and determined an average aspect ratio of 1.3 (ratio of width to height).  We find an average $A$ of 0.75$\pm$0.09 for both flux rope models, which when inverted corresponds to an aspect ratio of 1.33.

\subsection{FIDO Sensitivity}
Sections \ref{FIDO1} and \ref{BestFits} show that, for a small range in input parameters, FIDO can yield a range of quality-of-fit scores.  In this section we quantify FIDO's sensitivity to the flux rope latitude, longitude, and tilt by comparing the difference in these values between the four FIDO results for each ICME (ForeCAT-driven, GCS-driven, force free best fit, and circular best fit).

\begin{figure}[h!]
\centering
\includegraphics[width=\textwidth]{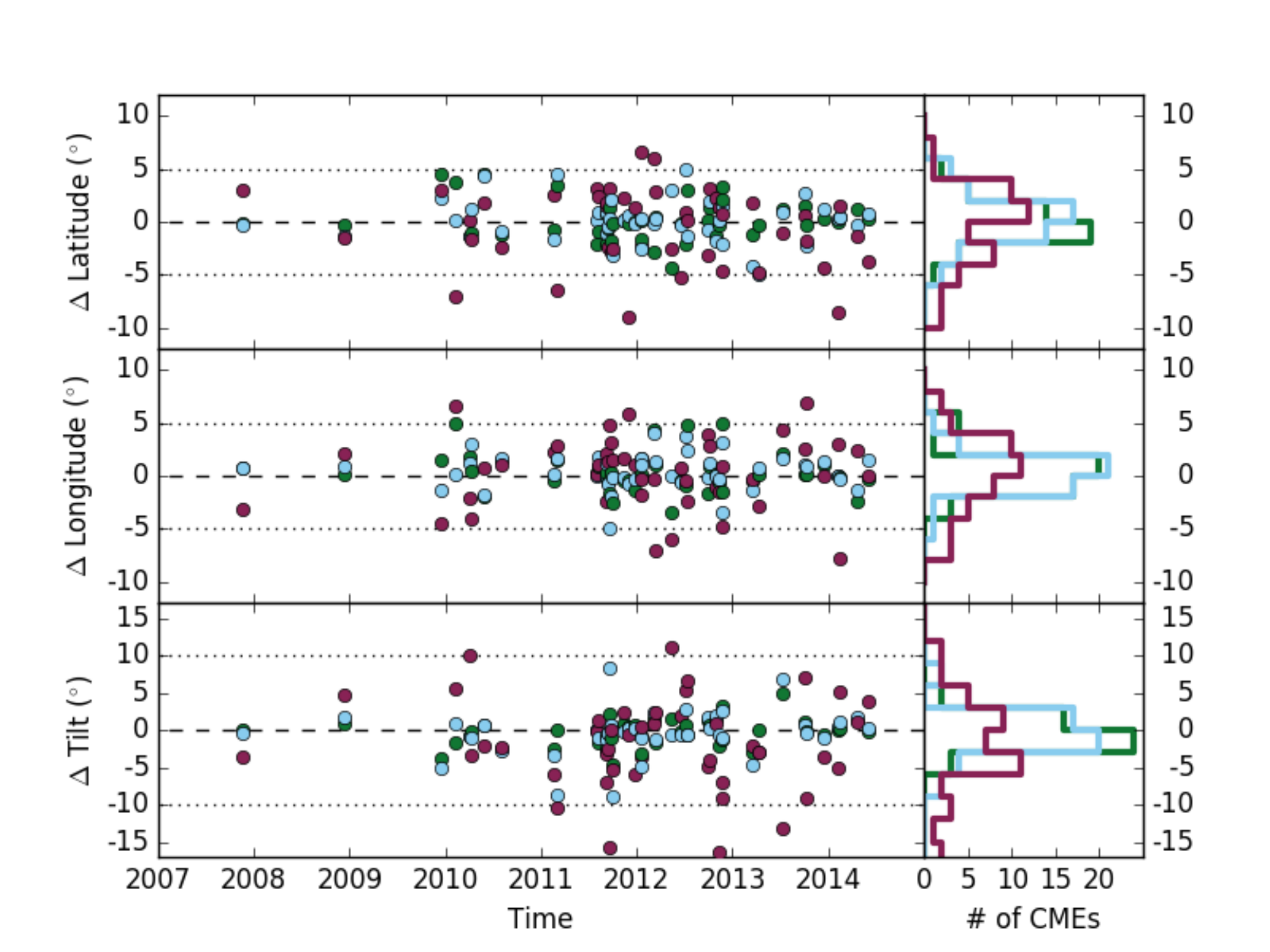}
\caption{Time series showing the difference in latitude (top), longitude (middle), and tilt (bottom) between the ForeCAT results and the force free best fit (green), the circular best fit (light blue), and the COR2 GCS results (purple).  The dotted lines show the maximum range allowed during the best fit parameter space exploration.  The right panels show histograms of the full time series.}
\label{fig:LLT}
\end{figure}

Figure \ref{fig:LLT} shows the difference from the ForeCAT-driven best fit for the GCS-driven (purple), force free best fit (green), and circular best fit (light blue).  The top panel shows the latitudinal difference with respect to time.  The middle and bottom panels show the longitudinal and orientation difference.  The dashed line at zero corresponds to no difference and the dotted lines within each panel show the range that was allowed during the best fit parameter space search.  The right panels collect the differences into histograms for the GCS-driven and two best fit results.

For most of the cases we find very small differences from the ForeCAT-driven results.  The best fit results consistently differ by less than the GCS-driven results.  For the force free and circular best fits we find average latitudinal differences of 1.4$^{\circ}\pm$1.3$^{\circ}$ and 1.5$^{\circ}\pm$1.4$^{\circ}$, longitudinal differences of 1.3$^{\circ}\pm$1.3$^{\circ}$ and 1.3$^{\circ}\pm$1.2$^{\circ}$, and tilt differences of 1.3$^{\circ}\pm$1.2$^{\circ}$ and 2.0$^{\circ}\pm$2.3$^{\circ}$, respectively.  In comparison, on average, the GCS differs by 3.0$^{\circ}\pm$2.0$^{\circ}$ in latitude, 2.6$^{\circ}\pm$2.0$^{\circ}$ in longitude, and 5.5$^{\circ}\pm$5.3$^{\circ}$ in tilt. 

As can be seen in Figure \ref{fig:LLT}, very few of the best fits have a latitude, longitude or tilt at the boundary of the sampled region. For the force free best fits, one ICME has a longitude differing by 5$^{\circ}$ (CME 4).  For the circular best fit we again find one varying by 5$^{\circ}$ in longitude (CME16).  Note that it is not the same cases showing the extreme variation between the two different flux rope models.  

This low number of extreme values has two implications.  First, we are considering a reasonable range when performing the best fit searches.  Our step size is such that the extreme values can be reached, if necessary, but the step size and range are appropriate for achieving convergence to a best fit within 4000 steps.  Second, the relatively small differences in the best fit and ForeCAT or GCS values confirm that little to no deflection or rotation occurs between the end of the ForeCAT simulation or coronal reconstruction and 1 AU.  Any deflection or rotation that occurs is smaller than the uncertainty in the GCS reconstructions.

\subsection{Impacts and Misses}
\citet{Kay17FIDO} suggest that near-Earth ICMEs can be divide into impacts and misses using two parameters derived from the ICME latitude and longitude relative to the Earth, the orientation, and the angular width.  We determine the angular distance between the Earth and the nose of the ICME, which we normalize using the angular width of the ICME.  When the normalized angular width is greater than one no impact should occur.  We also consider the proximity of the Earth to the toroidal axis of the ICME.  We expect that the ICME extends farther in the toroidal direction than in the poloidal direction/cross sectional width.  We quantify this distance using the angle between the toroidal axis and the line connecting the ICME nose and the Earth's position, which we normalize by 90$^{\circ}$ to put on the same scale as the normalized angular distance.  \citet{Kay17FIDO} show that ICMEs with smaller orientation differences tend to be more likely to impact the Earth.

\citet{Kay17FIDO} considered the results of over 600 different ForeCAT-driven FIDO simulations that represented four different observed cases.  Using all the simulations, \citet{Kay17FIDO} performed a linear regression to determine the line dividing the cases based on their normalized angular distance and orientation difference.  Here we compare that regression line, derived from only simulations and poorly constrained for high normalized angular widths, with the results for the 45 observed ICMEs considered in this paper.

\begin{figure}[h!]
\centering
\includegraphics[width=\textwidth]{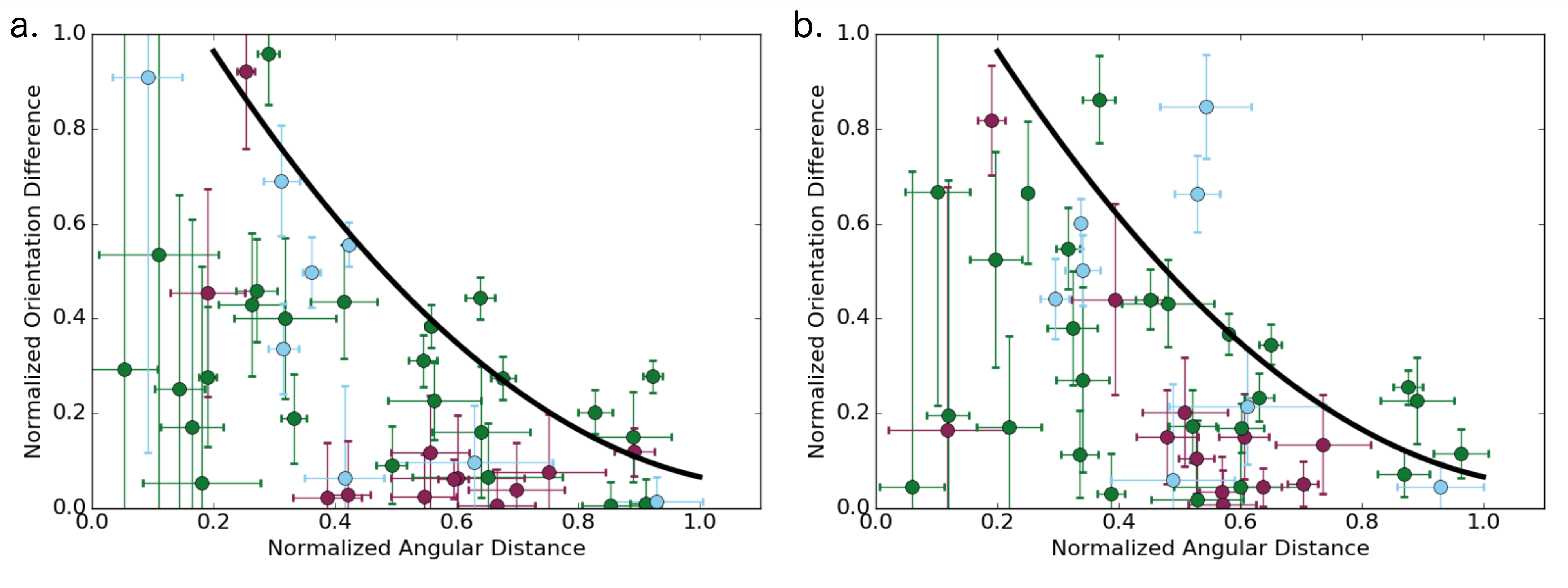}
\caption{The normalized difference of the clock angle between a CME's toroidal axis and the Earth's position versus the normalized angular distance of the Earth from the nose of the CME.  The color of the circle indicates the thickness of the CME according to their cross-sectional parameter $B$ with purple being thin ($B<0.25$), light blue being thick ($B>0.5$), and green falling in between the two extremes.  The error bars are derived by assuming uncertainties in latitude, longitude, and tilt corresponding to one quarter of the traditionally-assumed GCS values.  The black line shows the linear regression from \citet{Kay17FIDO}.  }
\label{fig:HorM}
\end{figure}

Figure \ref{fig:HorM} shows the normalized angular width and normalized orientation difference for the 45 ICMEs.  We estimate error bars for each case by assuming uncertainties of 1.25$^{\circ}$ in the latitude and 2.5$^{\circ}$ in the longitude, tilt, and angular width (equivalent to one quarter of the standard GCS values).  Note that these are not precisely determined error bars as setting the uncertainties at one quarter of the GCS uncertainties seems reasonable, but we have no justification for this specific value.  The error bars are simply meant to illustrate the sensitivity to uncertainty in the position, orientation, and angular width, and how it varies with the normalized angular distance and orientation difference.
  
Each circle is colored according to the thickness of the ICME using the shape parameter $B$, which describes the ratio of the cross-sectional width to the full width of the ICME.  Using the values from the ForeCAT-driven FIDO results, we color the thinnest ICMEs purple ($B<0.25$), the thickest light blue ($B>0.5$), and the in between green.  The black line shows the linear regression determined in \citet{Kay17FIDO}.  The left panel shows results using the ForeCAT position and orientation and the right panel shows results using the GCS reconstructions.

Since all 45 ICMEs impact Earth, we expect them to fall to the left and below the regression line in Figure \ref{fig:HorM}. The ForeCAT results in panel (a) show that the linear regression line works for most of these observed ICMEs.  Seven cases fall above the regression line (CMEs 1, 4, 10, 11, 13, 22, 25), but many of these have error bars that extend to or past the regression line, suggesting that a small change in their parameters could change them to a predicted hit.

We see that the cross-sectional thickness of the ICME influences where an ICME falls in Figure \ref{fig:HorM}.  In \citet{Kay17FIDO} the ICMEs all had $B$ betweeon 0.3 and 0.5 whereas the 45 ICMEs vary between 0.07 and 0.73.  As ICMEs become thicker, we expect that farther normalized distances will correspond to impacts for any given normalized orientation difference so that the regression line may shift up or to the right, depending on the actual sensitivity to $B$.  We find that the thinnest ICMEs tend to have very low normalized orientation differences, whereas the thickest ones have larger orientation differences. All the incorrectly predicted misses are medium or thick ICMEs, suggesting that the single line division, while reasonably accurate, could be improved by factoring in the ICME thickness.  This, however, is typically hard to measure remotely, so instead, conversion from a binary impact or miss designation to a probability of impact may be preferrable.

The right panel of Figure \ref{fig:HorM} shows the normalized angular distance and orientation difference for the GCS results.  Note that here the colors are determined using the $B$ used for the GCS-driven FIDO results and we use the same uncertainties as the left panel to determine the error bars.  We again see seven ICMEs would be incorrectly predicted misses, but these cases fall much farther to the right/above the linear regression line.  Again we see some structuring based upon the ICME's cross-sectional width, but with significantly more scatter than the ForeCAT results.  As seen in the in situ comparison, the GCS reconstruction tends to have a higher chance of large inaccuracies but typically tends to perform slightly worse than the ForeCAT-driven results.

\section{Discussion}\label{Disc}
In this paper we show that both the ForeCAT and GCS-reconstructed positions and orientations can be used with FIDO to yield acceptable fits to the in situ observations.  We wish to emphasize that neither the ForeCAT results nor the GCS reconstructions are fine-tuned best fits.  Both ForeCAT and the GCS model have degenericies between their input parameters.  For example, in ForeCAT a thicker cross section can often be compensated by a heavier mass, and a more inclined tilt may be balanced by a smaller angular width in the visual GCS fits.  For both, a minimum amount of parameter space searching was done to ensure that the CME does in fact impact the synthetic spacecraft and that corresponding FIDO results are reasonable.  Finding a set of parameters such that the ForeCAT and GCS results agree with one another and the FIDO results at least resemble the in situ observations often breaks the degeneracy between input parameters.  It is likely that better FIDO goodness-of-fit scores could be achieved by fine tuning both the ForeCAT and GCS models, but the goal of this work is to illustrate the capabilities of each model in a fashion that mimics how either may be used for real time predictions for a large number of cases, rather than detailed studies of a few select events.

These results show that, with a minimum amount of fine-tuning, the ForeCAT-driven FIDO tend to fit the in situ magnetic field than the GCS-driven FIDO results.  However, in a couple cases we find that the GCS-driven results actually perform better.  We are not advocating that the GCS reconstructions are inappropriate or inheritently inaccurate; they remain the best method of reconstructing CME positions and orientations, particularly when using multiple viewpoints.  We stress, however, that one must be careful when interpreting their results due to the large uncertainty in the parameters resulting from the impreciseness of visual fits and the degeneracy of the model parameters.

This paper and previous work \citep{Kay15L, Kay16Obs, Kay17AR, Cap17} have shown that ForeCAT can reproduce the deflection and rotation of observed CMEs, given the correct input parameters.  Future work will allow ForeCAT to be useful for actual predictions.  Given an AR where a CME may likely erupt from and the range of plausible CME parameters, ForeCAT runs efficiently enough to run large ensemble studies yielding probabilities of the possible of the possible deflections and rotations.  This range of positions and orientations, and their probabilites, could be coupled with FIDO to determine the range of magnetic profiles expected if the AR were to erupt and the ensuing CME were hit Earth.

\section{Conclusions}\label{Conc}
This paper presents the largest-to-date study of  observations and simulations of the deflection and rotation of CMEs and their magnetic profiles at 1 AU.  For 45 Earth-impacting CMEs between November 2007 and June 2014, we determine the initial location of each eruption, reconstruct their coronal position and orientation using the GCS model, simulate the coronal deflection and rotation using ForeCAT, and compare their magnetic field predicted by FIDO with in situ observations from ACE and Wind.  

The ForeCAT results and GCS reconstructions for the CME latitude, longitude, and tilt tend to agree within the uncertainty of the GCS reconstruction.  The combination of modeling and observations show that for these 45 CMEs, the largest deflections actually occur during solar minimum.  The deflection peaks during solar minimum due to the large latitudinal deflections at this time due to the smaller CME masses and enhanced polar magnetic field causing strong latitudinal gradients; the longitudinal deflections remain relatively constant over the full considered time span.  After the initial decrease early in the rising phase, the average deflections and rotations appear to be increasing as the solar cycle is approaching maximum, suggesting that the declining phase may have large deflections and rotations, but we cannot confirm this as the sample of CMEs does not extend that far.

Using the simulated deflection and rotation, we look for corrrelations between the magnitude of the deflection and rotation and various CME properties such as the mass, velocity, and angular width.  We recover many of the trends between CME properties known from large catalogs. The CME mass, angular width, and velocity are all positively correlated with one another. However, we only see weak correlations with the deflection and rotation, highlighting the complexity of the motion and the number of factors that simultaneously influence any single event.

We compare the in situ magnetic field with FIDO simulations driven by ForeCAT results.  For 42 of the 45 ICMEs we find a ``good'' fit, corresponding to a goodness-of-fit score less than one where one corresponds to assuming no magnetic field or twice the actual observed magnetic field for each of $B_x$, $B_y$, and $B_z$.  We find an average score of 0.72$\pm$0.22, which equals the average vector magnitude of the hourly error normalized by the total magnetic field strength.   The goodness-of-fit scores range between 0.24 and 1.05.  If we look at the individual components we find that all three vector components tend to have an hourly average error of 0.35$\pm$0.16, again normalized by the total ICME magnetic field strength.  We expect our simple flux rope model to work best for the ICMEs that appear the most flux-rope-like, and for the eight most flux-rope-like ICMEs (determined visually, though all are identified magnetic clouds) we find an average score of 0.55.  FIDO can still reproduce non-MC or ejecta-like events, which often correspond to flank hits, but less accurately than for the magnetic clouds.

In addition to using the ForeCAT results, we also drive the FIDO model using the GCS reconstructions.  These results range between 0.38 and 1.42 with an average score of 0.86$\pm$0.22, with eleven ICMEs having a score greater than one.  These results suggest that GCS reconstructions can be used to obtain good predictions of in situ magnetic field, however, there is a higher chance of poor predictions due to the larger uncertainties in the reconstructed latitude, longitude, and tilt.

To further explore the ability of ForeCAT to accurately determine the appropriate inputs for FIDO, we search for the best fit of the FIDO flux rope to the in situ observations.  This allows us to differentiate between the model's ability to reproduce the observations and our actual ability to determine the input parameters.  Considering both a force free and a circular flux rope model we find an average improvement of about 0.1 in the total goodness-of-fit score and about 0.05 in the individual vector components, with the circular model tending to yield marginally better fits.  These best fits tend to differ from the ForeCAT results by less than 2$^{\circ}$ in latitude and longitude and 3$^{\circ}$ in tilt.  This suggests that while ForeCAT yields good reconstructions, there is some room for improvement due to the extreme sensitivity to the ICME position and orientation.

Finally, we compare our simulated and reconstructed CME positions with the linear regression from \citet{Kay17FIDO} meant to divide CMEs into impacts and misses based upon their normalized angular width and normalized orientation difference.  We find that this criteria, developed from simulated CMEs, can be used to determine whether an observed CME will impact.  The results do show some sensitivity to the CME's width, so future work should incorporate a probability of impact instead of simply a binary impact or miss designation.

\acknowledgments
C.K.'s research was supported by an appointment to the NASA Postdoctoral Program at NASA GSFC, administered by the Universities Space Research Association under contract with NASA.  The authors thank T. Nieves-Chinchilla for useful information about the circular flux rope model.  The ACE data come from the Caltech Space Radiation Lab (http://www.srl.caltech.edu/ACE/ASC/level2/lvl2DATA\_MAG.html) and the Wind data from the NASA Goddard Space Flight Center Coordinated Data Analysis Web (http:cdaweb.sci.gsfc.nasa.gov).  The ICME start and stop times were taken from the ACE and Wind lists generously provided online at http://www.srl.caltech.edu/ACE/ASC/DATA/level3/icmetable2.htm and http://wind.nasa.gov/ICMEindex.php, and the CME velocity and masses from the SOHO LASCO CME catalog https://cdaw.gsfc.nasa.gov/CME\_list/.  The STEREO images, used for the GCS fits, came from https://stereo-ssc.nascom.nasa.gov/cgi-bin/images.

\listofchanges


\begin{thebibliography}{78}
\providecommand{\natexlab}[1]{#1}
\expandafter\ifx\csname urlstyle\endcsname\relax
  \providecommand{\doi}[1]{doi:\discretionary{}{}{}#1}\else
  \providecommand{\doi}{doi:\discretionary{}{}{}\begingroup
  \urlstyle{rm}\Url}\fi

\bibitem[{\textit{{Al-Haddad} et~al.}(2011)\textit{{Al-Haddad}, {Roussev},
  {M{\"o}stl}, {Jacobs}, {Lugaz}, {Poedts}, and {Farrugia}}}]{AlH11}
{Al-Haddad}, N., I.~I. {Roussev}, C.~{M{\"o}stl}, C.~{Jacobs}, N.~{Lugaz},
  S.~{Poedts}, and C.~J. {Farrugia} (2011), {On the Internal Structure of the
  Magnetic Field in Magnetic Clouds and Interplanetary Coronal Mass Ejections:
  Writhe versus Twist}, \textit{Astrophysical Journal Letters}, \textit{738}, L18,
  \doi{10.1088/2041-8205/738/2/L18}.

\bibitem[{\textit{{Berdichevsky}}(2013)}]{Ber13}
{Berdichevsky}, D.~B. (2013), {On Fields and Mass Constraints for the Uniform
  Propagation of Magnetic-Flux Ropes Undergoing Isotropic Expansion},
  \textit{Solar Physics}, \textit{284}, 245--259, \doi{10.1007/s11207-012-0176-5}.

\bibitem[{\textit{{Berdichevsky} et~al.}(2011)\textit{{Berdichevsky},
  {Stenborg}, and {Vourlidas}}}]{Ber11}
{Berdichevsky}, D.~B., G.~{Stenborg}, and A.~{Vourlidas} (2011), {Deriving the
  Physical Parameters of a Solar Ejection with an Isotropic Magnetohydrodynamic
  Evolutionary Model}, \textit{Astrophysical Journal }, \textit{741}, 47,
  \doi{10.1088/0004-637X/741/1/47}.

\bibitem[{\textit{{Bothmer} and {Schwenn}}(1994)}]{Bot94}
{Bothmer}, V., and R.~{Schwenn} (1994), {Eruptive prominences as sources of
  magnetic clouds in the solar wind}, \textit{Space Science Reviews}, \textit{70}, 215--220,
  \doi{10.1007/BF00777872}.

\bibitem[{\textit{{Bothmer} and {Schwenn}}(1998)}]{Bot98}
{Bothmer}, V., and R.~{Schwenn} (1998), {The structure and origin of magnetic
  clouds in the solar wind}, \textit{Annales Geophysicae}, \textit{16}, 1--24,
  \doi{10.1007/s00585-997-0001-x}.

\bibitem[{\textit{{Burlaga} et~al.}(1981)\textit{{Burlaga}, {Sittler},
  {Mariani}, and {Schwenn}}}]{Bur81}
{Burlaga}, L., E.~{Sittler}, F.~{Mariani}, and R.~{Schwenn} (1981), {Magnetic
  loop behind an interplanetary shock - Voyager, Helios, and IMP 8
  observations}, \textit{Journal of Geophysical Research}, \textit{86}, 6673--6684,
  \doi{10.1029/JA086iA08p06673}.

\bibitem[{\textit{{Byrne} et~al.}(2010)\textit{{Byrne}, {Maloney}, {McAteer},
  {Refojo}, and {Gallagher}}}]{Byr10}
{Byrne}, J.~P., S.~A. {Maloney}, R.~T.~J. {McAteer}, J.~M. {Refojo}, and P.~T.
  {Gallagher} (2010), {Propagation of an Earth-directed coronal mass ejection
  in three dimensions}, \textit{Nature Communications}, \textit{1}, 74,
  \doi{10.1038/ncomms1077}.

\bibitem[{\textit{{Capannolo} et~al.}(2017)\textit{{Capannolo}, {Opher}, {Kay},
  and {Landi}}}]{Cap17}
{Capannolo}, L., M.~{Opher}, C.~{Kay}, and E.~{Landi} (2017), {The Deflection
  of the Cartwheel CME: ForeCAT Results}, \textit{Astrophysical Journal }, \textit{839}, 37,
  \doi{10.3847/1538-4357/aa6a16}.

\bibitem[{\textit{{Cremades} and {Bothmer}}(2004)}]{Cre04}
{Cremades}, H., and V.~{Bothmer} (2004), {On the three-dimensional
  configuration of coronal mass ejections}, \textit{Astron. and Astrophys.}, \textit{422},
  307--322, \doi{10.1051/0004-6361:20035776}.

\bibitem[{\textit{{Farrugia} et~al.}(1993)\textit{{Farrugia}, {Burlaga},
  {Osherovich}, {Richardson}, {Freeman}, {Lepping}, and {Lazarus}}}]{Far93}
{Farrugia}, C.~J., L.~F. {Burlaga}, V.~A. {Osherovich}, I.~G. {Richardson},
  M.~P. {Freeman}, R.~P. {Lepping}, and A.~J. {Lazarus} (1993), {A study of an
  expanding interplanatary magnetic cloud and its interaction with the earth's
  magnetosphere - The interplanetary aspect}, \textit{Journal of Geophysical Research}, \textit{98},
  7621--7632, \doi{10.1029/92JA02349}.

\bibitem[{\textit{{Filippov} et~al.}(2001)\textit{{Filippov}, {Gopalswamy}, and
  {Lozhechkin}}}]{Fil01}
{Filippov}, B.~P., N.~{Gopalswamy}, and A.~V. {Lozhechkin} (2001), {Non-radial
  motion of eruptive filaments}, \textit{Solar Physics}, \textit{203}, 119--130,
  \doi{10.1023/A:1012754329767}.

\bibitem[{\textit{{Gopalswamy}}(2008)}]{Gop08Geo}
{Gopalswamy}, N. (2008), {Solar connections of geoeffective magnetic
  structures}, \textit{Journal of Atmospheric and Solar-Terrestrial Physics},
  \textit{70}, 2078--2100, \doi{10.1016/j.jastp.2008.06.010}.

\bibitem[{\textit{{Gopalswamy}}(2015)}]{Gop15EP}
{Gopalswamy}, N. (2015), {The Dynamics of Eruptive Prominences}, in
  \textit{Solar Prominences}, \textit{Astrophysics and Space Science Library},
  vol. 415, edited by J.-C. {Vial} and O.~{Engvold}, p. 381,
  \doi{10.1007/978-3-319-10416-4\_15}.

\bibitem[{\textit{{Gopalswamy} et~al.}(2008)\textit{{Gopalswamy}, {Akiyama},
  {Yashiro}, {Michalek}, and {Lepping}}}]{Gop08}
{Gopalswamy}, N., S.~{Akiyama}, S.~{Yashiro}, G.~{Michalek}, and R.~P.
  {Lepping} (2008), {Solar sources and geospace consequences of interplanetary
  magnetic clouds observed during solar cycle 23}, \textit{Journal of
  Atmospheric and Solar-Terrestrial Physics}, \textit{70}, 245--253,
  \doi{10.1016/j.jastp.2007.08.070}.

\bibitem[{\textit{{Gopalswamy} et~al.}(2009)\textit{{Gopalswamy},
  {M{\"a}kel{\"a}}, {Xie}, {Akiyama}, and {Yashiro}}}]{Gop09}
{Gopalswamy}, N., P.~{M{\"a}kel{\"a}}, H.~{Xie}, S.~{Akiyama}, and S.~{Yashiro}
  (2009), {CME interactions with coronal holes and their interplanetary
  consequences}, \textit{Journal of Geophysical Research (Space Physics)},
  \textit{114}, A00A22, \doi{10.1029/2008JA013686}.

\bibitem[{\textit{{Gopalswamy} et~al.}(2013)\textit{{Gopalswamy},
  {M{\"a}kel{\"a}}, {Xie}, and {Yashiro}}}]{Gop13}
{Gopalswamy}, N., P.~{M{\"a}kel{\"a}}, H.~{Xie}, and S.~{Yashiro} (2013),
  {Testing the empirical shock arrival model using quadrature observations},
  \textit{Space Weather}, \textit{11}, 661--669, \doi{10.1002/2013SW000945}.

\bibitem[{\textit{{Gopalswamy} et~al.}(2017)\textit{{Gopalswamy}, {Akiyama},
  {Yashiro}, and {Xie}}}]{Gop17b}
{Gopalswamy}, N., S.~{Akiyama}, S.~{Yashiro}, and H.~{Xie} (2017), {Coronal
  Flux Ropes and their Interplanetary Counterparts}, \textit{ArXiv e-prints}.

\bibitem[{\textit{{Green} et~al.}(2007)\textit{{Green}, {Kliem},
  {T{\"o}r{\"o}k}, {van Driel-Gesztelyi}, and {Attrill}}}]{Gre07}
{Green}, L.~M., B.~{Kliem}, T.~{T{\"o}r{\"o}k}, L.~{van Driel-Gesztelyi}, and
  G.~D.~R. {Attrill} (2007), {Transient Coronal Sigmoids and Rotating Erupting
  Flux Ropes}, \textit{Solar Physics}, \textit{246}, 365--391,
  \doi{10.1007/s11207-007-9061-z}.

\bibitem[{\textit{{Gui} et~al.}(2011)\textit{{Gui}, {Shen}, {Wang}, {Ye},
  {Liu}, {Wang}, and {Zhao}}}]{Gui11}
{Gui}, B., C.~{Shen}, Y.~{Wang}, P.~{Ye}, J.~{Liu}, S.~{Wang}, and X.~{Zhao}
  (2011), {Quantitative Analysis of CME Deflections in the Corona},
  \textit{Solar Physics}, \textit{271}, 111--139, \doi{10.1007/s11207-011-9791-9}.

\bibitem[{\textit{{Hildner}}(1977)}]{Hil77}
{Hildner}, E. (1977), {Mass ejections from the solar corona into interplanetary
  space}, in \textit{Study of Travelling Interplanetary Phenomena},
  \textit{Astrophysics and Space Science Library}, vol.~71, edited by M.~A.
  {Shea}, D.~F. {Smart}, and S.~T. {Wu}, pp. 3--20.

\bibitem[{\textit{{Hu} and {Sonnerup}}(2002)}]{Hu02}
{Hu}, Q., and B.~U.~{\"O}. {Sonnerup} (2002), {Reconstruction of magnetic
  clouds in the solar wind: Orientations and configurations}, \textit{Journal
  of Geophysical Research (Space Physics)}, \textit{107}, 1142,
  \doi{10.1029/2001JA000293}.

\bibitem[{\textit{{Isavnin}}(2016)}]{Isa16}
{Isavnin}, A. (2016), {FRiED: A Novel Three-dimensional Model of Coronal Mass
  Ejections}, \textit{Astrophysical Journal }, \textit{833}, 267,
  \doi{10.3847/1538-4357/833/2/267}.

\bibitem[{\textit{{Isavnin} et~al.}(2013)\textit{{Isavnin}, {Vourlidas}, and
  {Kilpua}}}]{Isa13}
{Isavnin}, A., A.~{Vourlidas}, and E.~K.~J. {Kilpua} (2013), {Three-Dimensional
  Evolution of Erupted Flux Ropes from the Sun (2 - 20 Rs) to 1 AU},
  \textit{Solar Physics}, \textit{284}, 203--215, \doi{10.1007/s11207-012-0214-3}.

\bibitem[{\textit{{Isavnin} et~al.}(2014)\textit{{Isavnin}, {Vourlidas}, and
  {Kilpua}}}]{Isa14}
{Isavnin}, A., A.~{Vourlidas}, and E.~K.~J. {Kilpua} (2014), {Three-Dimensional
  Evolution of Flux-Rope CMEs and Its Relation to the Local Orientation of the
  Heliospheric Current Sheet}, \textit{Solar Physics}, \textit{289}, 2141--2156,
  \doi{10.1007/s11207-013-0468-4}.

\bibitem[{\textit{{Jakosky} et~al.}(2015)\textit{{Jakosky}, {Grebowsky},
  {Luhmann}, {Connerney}, {Eparvier}, {Ergun}, {Halekas}, {Larson}, {Mahaffy},
  {McFadden}, {Mitchell}, {Schneider}, {Zurek}, {Bougher}, {Brain}, {Ma},
  {Mazelle}, {Andersson}, {Andrews}, {Baird}, {Baker}, {Bell}, {Benna},
  {Chaffin}, {Chamberlin}, {Chaufray}, {Clarke}, {Collinson}, {Combi}, {Crary},
  {Cravens}, {Crismani}, {Curry}, {Curtis}, {Deighan}, {Delory}, {Dewey},
  {DiBraccio}, {Dong}, {Dong}, {Dunn}, {Elrod}, {England}, {Eriksson},
  {Espley}, {Evans}, {Fang}, {Fillingim}, {Fortier}, {Fowler}, {Fox},
  {Gr{\"o}ller}, {Guzewich}, {Hara}, {Harada}, {Holsclaw}, {Jain}, {Jolitz},
  {Leblanc}, {Lee}, {Lee}, {Lefevre}, {Lillis}, {Livi}, {Lo}, {Mayyasi},
  {McClintock}, {McEnulty}, {Modolo}, {Montmessin}, {Morooka}, {Nagy}, {Olsen},
  {Peterson}, {Rahmati}, {Ruhunusiri}, {Russell}, {Sakai}, {Sauvaud}, {Seki},
  {Steckiewicz}, {Stevens}, {Stewart}, {Stiepen}, {Stone}, {Tenishev},
  {Thiemann}, {Tolson}, {Toublanc}, {Vogt}, {Weber}, {Withers}, {Woods}, and
  {Yelle}}}]{Jak15}
{Jakosky}, B.~M., J.~M. {Grebowsky}, J.~G. {Luhmann}, J.~{Connerney},
  F.~{Eparvier}, R.~{Ergun}, J.~{Halekas}, D.~{Larson}, P.~{Mahaffy},
  J.~{McFadden}, D.~F. {Mitchell}, N.~{Schneider}, R.~{Zurek}, S.~{Bougher},
  D.~{Brain}, Y.~J. {Ma}, C.~{Mazelle}, L.~{Andersson}, D.~{Andrews},
  D.~{Baird}, D.~{Baker}, J.~M. {Bell}, M.~{Benna}, M.~{Chaffin},
  P.~{Chamberlin}, Y.-Y. {Chaufray}, J.~{Clarke}, G.~{Collinson}, M.~{Combi},
  F.~{Crary}, T.~{Cravens}, M.~{Crismani}, S.~{Curry}, D.~{Curtis},
  J.~{Deighan}, G.~{Delory}, R.~{Dewey}, G.~{DiBraccio}, C.~{Dong}, Y.~{Dong},
  P.~{Dunn}, M.~{Elrod}, S.~{England}, A.~{Eriksson}, J.~{Espley}, S.~{Evans},
  X.~{Fang}, M.~{Fillingim}, K.~{Fortier}, C.~M. {Fowler}, J.~{Fox},
  H.~{Gr{\"o}ller}, S.~{Guzewich}, T.~{Hara}, Y.~{Harada}, G.~{Holsclaw}, S.~K.
  {Jain}, R.~{Jolitz}, F.~{Leblanc}, C.~O. {Lee}, Y.~{Lee}, F.~{Lefevre},
  R.~{Lillis}, R.~{Livi}, D.~{Lo}, M.~{Mayyasi}, W.~{McClintock},
  T.~{McEnulty}, R.~{Modolo}, F.~{Montmessin}, M.~{Morooka}, A.~{Nagy},
  K.~{Olsen}, W.~{Peterson}, A.~{Rahmati}, S.~{Ruhunusiri}, C.~T. {Russell},
  S.~{Sakai}, J.-A. {Sauvaud}, K.~{Seki}, M.~{Steckiewicz}, M.~{Stevens},
  A.~I.~F. {Stewart}, A.~{Stiepen}, S.~{Stone}, V.~{Tenishev}, E.~{Thiemann},
  R.~{Tolson}, D.~{Toublanc}, M.~{Vogt}, T.~{Weber}, P.~{Withers}, T.~{Woods},
  and R.~{Yelle} (2015), {MAVEN observations of the response of Mars to an
  interplanetary coronal mass ejection}, \textit{Science}, \textit{350}, 0210,
  \doi{10.1126/science.aad0210}.

\bibitem[{\textit{{Janvier} et~al.}(2013)\textit{{Janvier}, {D{\'e}moulin}, and
  {Dasso}}}]{Jan13}
{Janvier}, M., P.~{D{\'e}moulin}, and S.~{Dasso} (2013), {Global axis shape of
  magnetic clouds deduced from the distribution of their local axis
  orientation}, \textit{Astron. and Astrophys.}, \textit{556}, A50,
  \doi{10.1051/0004-6361/201321442}.

\bibitem[{\textit{{Janvier} et~al.}(2015)\textit{{Janvier}, {Dasso},
  {D{\'e}moulin}, {Mas{\'{\i}}as-Meza}, and {Lugaz}}}]{Jan15}
{Janvier}, M., S.~{Dasso}, P.~{D{\'e}moulin}, J.~J. {Mas{\'{\i}}as-Meza}, and
  N.~{Lugaz} (2015), {Comparing generic models for interplanetary shocks and
  magnetic clouds axis configurations at 1 AU}, \textit{Journal of Geophysical
  Research (Space Physics)}, \textit{120}, 3328--3349,
  \doi{10.1002/2014JA020836}.

\bibitem[{\textit{{Jin} et~al.}(2017)\textit{{Jin}, {Manchester}, {van der
  Holst}, {Sokolov}, {T{\'o}th}, {Vourlidas}, {de Koning}, and
  {Gombosi}}}]{Jin17}
{Jin}, M., W.~B. {Manchester}, B.~{van der Holst}, I.~{Sokolov}, G.~{T{\'o}th},
  A.~{Vourlidas}, C.~A. {de Koning}, and T.~I. {Gombosi} (2017), {Chromosphere
  to 1 AU Simulation of the 2011 March 7th Event: A Comprehensive Study of
  Coronal Mass Ejection Propagation}, \textit{Astrophysical Journal }, \textit{834}, 172,
  \doi{10.3847/1538-4357/834/2/172}.

\bibitem[{\textit{{Kay} and {Opher}}(2015)}]{Kay15AM}
{Kay}, C., and M.~{Opher} (2015), {The Heliocentric Distance where the
  Deflections and Rotations of Solar Coronal Mass Ejections Occur},
  \textit{Astrophysical Journal Letters}, \textit{811}, L36, \doi{10.1088/2041-8205/811/2/L36}.

\bibitem[{\textit{{Kay} et~al.}(2013)\textit{{Kay}, {Opher}, and
  {Evans}}}]{Kay13}
{Kay}, C., M.~{Opher}, and R.~M. {Evans} (2013), {Forecasting a Coronal Mass
  Ejection's Altered Trajectory: ForeCAT}, \textit{Astrophysical Journal }, \textit{775}, 5,
  \doi{10.1088/0004-637X/775/1/5}.

\bibitem[{\textit{{Kay} et~al.}(2015{\natexlab{a}})\textit{{Kay}, {Opher}, and
  {Evans}}}]{Kay15}
{Kay}, C., M.~{Opher}, and R.~M. {Evans} (2015{\natexlab{a}}), {Global Trends
  of CME Deflections Based on CME and Solar Parameters}, \textit{Astrophysical Journal },
  \textit{805}, 168, \doi{10.1088/0004-637X/805/2/168}.

\bibitem[{\textit{{Kay} et~al.}(2015{\natexlab{b}})\textit{{Kay}, {dos Santos},
  and {Opher}}}]{Kay15L}
{Kay}, C., L.~F.~G. {dos Santos}, and M.~{Opher} (2015{\natexlab{b}}),
  {Constraining the Masses and the Non-radial Drag Coefficient of a Solar
  Coronal Mass Ejection}, \textit{Astrophysical Journal Letters}, \textit{801}, L21,
  \doi{10.1088/2041-8205/801/2/L21}.

\bibitem[{\textit{{Kay} et~al.}(2016)\textit{{Kay}, {Opher}, {Colaninno}, and
  {Vourlidas}}}]{Kay16Obs}
{Kay}, C., M.~{Opher}, R.~C. {Colaninno}, and A.~{Vourlidas} (2016), {Using
  ForeCAT Deflections and Rotations to Constrain the Early Evolution of CMEs},
  \textit{Astrophysical Journal }, \textit{827}, 70, \doi{10.3847/0004-637X/827/1/70}.

\bibitem[{\textit{{Kay} et~al.}(2017{\natexlab{a}})\textit{{Kay}, {Gopalswamy},
  {Reinard}, and {Opher}}}]{Kay17FIDO}
{Kay}, C., N.~{Gopalswamy}, A.~{Reinard}, and M.~{Opher} (2017{\natexlab{a}}),
  {Predicting the Magnetic Field of Earth-impacting CMEs}, \textit{Astrophysical Journal },
  \textit{835}, 117, \doi{10.3847/1538-4357/835/2/117}.

\bibitem[{\textit{{Kay} et~al.}(2017{\natexlab{b}})\textit{{Kay}, {Gopalswamy},
  {Xie}, and {Yashiro}}}]{Kay17AR}
{Kay}, C., N.~{Gopalswamy}, H.~{Xie}, and S.~{Yashiro} (2017{\natexlab{b}}),
  {Deflection and Rotation of CMEs from Active Region 11158},
  \textit{Solar Physics}, \textit{292}, 78, \doi{10.1007/s11207-017-1098-z}.

\bibitem[{\textit{{Kilpua} et~al.}(2009)\textit{{Kilpua}, {Pomoell},
  {Vourlidas}, {Vainio}, {Luhmann}, {Li}, {Schroeder}, {Galvin}, and
  {Simunac}}}]{Kil09}
{Kilpua}, E.~K.~J., J.~{Pomoell}, A.~{Vourlidas}, R.~{Vainio}, J.~{Luhmann},
  Y.~{Li}, P.~{Schroeder}, A.~B. {Galvin}, and K.~{Simunac} (2009), {STEREO
  observations of interplanetary coronal mass ejections and prominence
  deflection during solar minimum period}, \textit{Annales Geophysicae},
  \textit{27}, 4491--4503, \doi{10.5194/angeo-27-4491-2009}.

\bibitem[{\textit{{Kliem} et~al.}(2012)\textit{{Kliem}, {T{\"o}r{\"o}k}, and
  {Thompson}}}]{Kli12}
{Kliem}, B., T.~{T{\"o}r{\"o}k}, and W.~T. {Thompson} (2012), {A Parametric
  Study of Erupting Flux Rope Rotation. Modeling the ''Cartwheel CME'' on 9
  April 2008}, \textit{Solar Physics}, \textit{281}, 137--166,
  \doi{10.1007/s11207-012-9990-z}.

\bibitem[{\textit{{Kunkel} and {Chen}}(2010)}]{Kun10}
{Kunkel}, V., and J.~{Chen} (2010), {Evolution of a Coronal Mass Ejection and
  its Magnetic Field in Interplanetary Space}, \textit{Astrophysical Journal Letters}, \textit{715},
  L80--L83, \doi{10.1088/2041-8205/715/2/L80}.

\bibitem[{\textit{{Lepping} et~al.}(1990)\textit{{Lepping}, {Burlaga}, and
  {Jones}}}]{Lep90}
{Lepping}, R.~P., L.~F. {Burlaga}, and J.~A. {Jones} (1990), {Magnetic field
  structure of interplanetary magnetic clouds at 1 AU}, \textit{Journal of Geophysical Research},
  \textit{95}, 11,957--11,965, \doi{10.1029/JA095iA08p11957}.

\bibitem[{\textit{{Lepping} et~al.}(2011)\textit{{Lepping}, {Wu},
  {Berdichevsky}, and {Szabo}}}]{Lep11}
{Lepping}, R.~P., C.-C. {Wu}, D.~B. {Berdichevsky}, and A.~{Szabo} (2011),
  {Magnetic Clouds at/near the 2007 - 2009 Solar Minimum: Frequency of
  Occurrence and Some Unusual Properties}, \textit{Solar Physics}, \textit{274},
  345--360, \doi{10.1007/s11207-010-9646-9}.

\bibitem[{\textit{{Lepping} et~al.}(2015)\textit{{Lepping}, {Wu},
  {Berdichevsky}, and {Szabo}}}]{Lep15}
{Lepping}, R.~P., C.-C. {Wu}, D.~B. {Berdichevsky}, and A.~{Szabo} (2015),
  {Wind Magnetic Clouds for 2010 - 2012: Model Parameter Fittings, Associated
  Shock Waves, and Comparisons to Earlier Periods}, \textit{Solar Physics},
  \textit{290}, 2265--2290, \doi{10.1007/s11207-015-0755-3}.

\bibitem[{\textit{{Liu} et~al.}(2010)\textit{{Liu}, {Davies}, {Luhmann},
  {Vourlidas}, {Bale}, and {Lin}}}]{Liu10a}
{Liu}, Y., J.~A. {Davies}, J.~G. {Luhmann}, A.~{Vourlidas}, S.~D. {Bale}, and
  R.~P. {Lin} (2010), {Geometric Triangulation of Imaging Observations to Track
  Coronal Mass Ejections Continuously Out to 1 AU}, \textit{Astrophysical Journal Letters},
  \textit{710}, L82--L87, \doi{10.1088/2041-8205/710/1/L82}.

\bibitem[{\textit{{Lugaz} et~al.}(2010)\textit{{Lugaz}, {Hernandez-Charpak},
  {Roussev}, {Davis}, {Vourlidas}, and {Davies}}}]{Lug10}
{Lugaz}, N., J.~N. {Hernandez-Charpak}, I.~I. {Roussev}, C.~J. {Davis},
  A.~{Vourlidas}, and J.~A. {Davies} (2010), {Determining the Azimuthal
  Properties of Coronal Mass Ejections from Multi-Spacecraft Remote-Sensing
  Observations with STEREO SECCHI}, \textit{Astrophysical Journal }, \textit{715}, 493--499,
  \doi{10.1088/0004-637X/715/1/493}.

\bibitem[{\textit{{Lugaz} et~al.}(2012)\textit{{Lugaz}, {Farrugia}, {Davies},
  {M{\"o}stl}, {Davis}, {Roussev}, and {Temmer}}}]{Lug12}
{Lugaz}, N., C.~J. {Farrugia}, J.~A. {Davies}, C.~{M{\"o}stl}, C.~J. {Davis},
  I.~I. {Roussev}, and M.~{Temmer} (2012), {The Deflection of the Two
  Interacting Coronal Mass Ejections of 2010 May 23-24 as Revealed by Combined
  in Situ Measurements and Heliospheric Imaging}, \textit{Astrophysical Journal }, \textit{759},
  68, \doi{10.1088/0004-637X/759/1/68}.

\bibitem[{\textit{{Lugaz} et~al.}(2013)\textit{{Lugaz}, {Farrugia},
  {Manchester}, and {Schwadron}}}]{Lug13}
{Lugaz}, N., C.~J. {Farrugia}, W.~B. {Manchester}, IV, and N.~{Schwadron}
  (2013), {The Interaction of Two Coronal Mass Ejections: Influence of Relative
  Orientation}, \textit{Astrophysical Journal }, \textit{778}, 20,
  \doi{10.1088/0004-637X/778/1/20}.

\bibitem[{\textit{{Lynch} et~al.}(2009)\textit{{Lynch}, {Antiochos}, {Li},
  {Luhmann}, and {DeVore}}}]{Lyn09}
{Lynch}, B.~J., S.~K. {Antiochos}, Y.~{Li}, J.~G. {Luhmann}, and C.~R. {DeVore}
  (2009), {Rotation of Coronal Mass Ejections during Eruption}, \textit{Astrophysical Journal },
  \textit{697}, 1918--1927, \doi{10.1088/0004-637X/697/2/1918}.

\bibitem[{\textit{{MacQueen} et~al.}(1986)\textit{{MacQueen}, {Hundhausen}, and
  {Conover}}}]{Mac86}
{MacQueen}, R.~M., A.~J. {Hundhausen}, and C.~W. {Conover} (1986), {The
  propagation of coronal mass ejection transients}, \textit{Journal of Geophysical Research}, \textit{91},
  31--38, \doi{10.1029/JA091iA01p00031}.

\bibitem[{\textit{{M{\"o}stl} et~al.}(2008)\textit{{M{\"o}stl}, {Miklenic},
  {Farrugia}, {Temmer}, {Veronig}, {Galvin}, {Vr{\v s}nak}, and
  {Biernat}}}]{Mos08}
{M{\"o}stl}, C., C.~{Miklenic}, C.~J. {Farrugia}, M.~{Temmer}, A.~{Veronig},
  A.~B. {Galvin}, B.~{Vr{\v s}nak}, and H.~K. {Biernat} (2008), {Two-spacecraft
  reconstruction of a magnetic cloud and comparison to its solar source},
  \textit{Annales Geophysicae}, \textit{26}, 3139--3152,
  \doi{10.5194/angeo-26-3139-2008}.

\bibitem[{\textit{{M{\"o}stl} et~al.}(2009)\textit{{M{\"o}stl}, {Farrugia},
  {Temmer}, {Miklenic}, {Veronig}, {Galvin}, {Leitner}, and {Biernat}}}]{Mos09}
{M{\"o}stl}, C., C.~J. {Farrugia}, M.~{Temmer}, C.~{Miklenic}, A.~M. {Veronig},
  A.~B. {Galvin}, M.~{Leitner}, and H.~K. {Biernat} (2009), {Linking Remote
  Imagery of a Coronal Mass Ejection to Its In Situ Signatures at 1 AU},
  \textit{Astrophysical Journal Letters}, \textit{705}, L180--L185, \doi{10.1088/0004-637X/705/2/L180}.

\bibitem[{\textit{{M{\"o}stl} et~al.}(2014)\textit{{M{\"o}stl}, {Amla}, {Hall},
  {Liewer}, {De Jong}, {Colaninno}, {Veronig}, {Rollett}, {Temmer}, {Peinhart},
  {Davies}, {Lugaz}, {Liu}, {Farrugia}, {Luhmann}, {Vr{\v s}nak}, {Harrison},
  and {Galvin}}}]{Mos14}
{M{\"o}stl}, C., K.~{Amla}, J.~R. {Hall}, P.~C. {Liewer}, E.~M. {De Jong},
  R.~C. {Colaninno}, A.~M. {Veronig}, T.~{Rollett}, M.~{Temmer}, V.~{Peinhart},
  J.~A. {Davies}, N.~{Lugaz}, Y.~D. {Liu}, C.~J. {Farrugia}, J.~G. {Luhmann},
  B.~{Vr{\v s}nak}, R.~A. {Harrison}, and A.~B. {Galvin} (2014), {Connecting
  Speeds, Directions and Arrival Times of 22 Coronal Mass Ejections from the
  Sun to 1 AU}, \textit{Astrophysical Journal }, \textit{787}, 119,
  \doi{10.1088/0004-637X/787/2/119}.

\bibitem[{\textit{{M{\"o}stl} et~al.}(2015)\textit{{M{\"o}stl}, {Rollett},
  {Frahm}, {Liu}, {Long}, {Colaninno}, {Reiss}, {Temmer}, {Farrugia}, {Posner},
  {Dumbovi{\'c}}, {Janvier}, {D{\'e}moulin}, {Boakes}, {Devos}, {Kraaikamp},
  {Mays}, and {Vr{\v s}nak}}}]{Mos15}
{M{\"o}stl}, C., T.~{Rollett}, R.~A. {Frahm}, Y.~D. {Liu}, D.~M. {Long}, R.~C.
  {Colaninno}, M.~A. {Reiss}, M.~{Temmer}, C.~J. {Farrugia}, A.~{Posner},
  M.~{Dumbovi{\'c}}, M.~{Janvier}, P.~{D{\'e}moulin}, P.~{Boakes}, A.~{Devos},
  E.~{Kraaikamp}, M.~L. {Mays}, and B.~{Vr{\v s}nak} (2015), {Strong coronal
  channelling and interplanetary evolution of a solar storm up to Earth and
  Mars}, \textit{Nature Communications}, \textit{6}, 7135,
  \doi{10.1038/ncomms8135}.

\bibitem[{\textit{{Nieves-Chinchilla} et~al.}(2013)\textit{{Nieves-Chinchilla},
  {Vourlidas}, {Stenborg}, {Savani}, {Koval}, {Szabo}, and {Jian}}}]{Nie13}
{Nieves-Chinchilla}, T., A.~{Vourlidas}, G.~{Stenborg}, N.~P. {Savani},
  A.~{Koval}, A.~{Szabo}, and L.~K. {Jian} (2013), {Inner Heliospheric
  Evolution of a ''Stealth'' CME Derived from Multi-view Imaging and Multipoint
  in Situ observations. I. Propagation to 1 AU}, \textit{Astrophysical Journal }, \textit{779},
  55, \doi{10.1088/0004-637X/779/1/55}.

\bibitem[{\textit{{Nieves-Chinchilla} et~al.}(2016)\textit{{Nieves-Chinchilla},
  {Linton}, {Hidalgo}, {Vourlidas}, {Savani}, {Szabo}, {Farrugia}, and
  {Yu}}}]{Nie16}
{Nieves-Chinchilla}, T., M.~G. {Linton}, M.~A. {Hidalgo}, A.~{Vourlidas}, N.~P.
  {Savani}, A.~{Szabo}, C.~{Farrugia}, and W.~{Yu} (2016), {A
  Circular-cylindrical Flux-rope Analytical Model for Magnetic Clouds},
  \textit{Astrophysical Journal }, \textit{823}, 27, \doi{10.3847/0004-637X/823/1/27}.

\bibitem[{\textit{{Owens}}(2008)}]{Owe08}
{Owens}, M.~J. (2008), {Combining remote and in situ observations of coronal
  mass ejections to better constrain magnetic cloud reconstruction},
  \textit{Journal of Geophysical Research (Space Physics)}, \textit{113},
  A12102, \doi{10.1029/2008JA013589}.

\bibitem[{\textit{{Palmerio} et~al.}(2017)\textit{{Palmerio}, {Kilpua},
  {James}, {Green}, {Pomoell}, {Isavnin}, and {Valori}}}]{Pal17}
{Palmerio}, E., E.~K.~J. {Kilpua}, A.~W. {James}, L.~M. {Green}, J.~{Pomoell},
  A.~{Isavnin}, and G.~{Valori} (2017), {Determining the Intrinsic CME Flux
  Rope Type Using Remote-sensing Solar Disk Observations}, \textit{Solar Physics},
  \textit{292}, 39, \doi{10.1007/s11207-017-1063-x}.

\bibitem[{\textit{{Pevtsov} et~al.}(2014)\textit{{Pevtsov}, {Berger}, {Nindos},
  {Norton}, and {van Driel-Gesztelyi}}}]{Pet14}
{Pevtsov}, A.~A., M.~A. {Berger}, A.~{Nindos}, A.~A. {Norton}, and L.~{van
  Driel-Gesztelyi} (2014), {Magnetic Helicity, Tilt, and Twist}, \textit{Space Science Reviews},
  \textit{186}, 285--324, \doi{10.1007/s11214-014-0082-2}.

\bibitem[{\textit{{Richardson} and {Cane}}(2010)}]{Ric10}
{Richardson}, I.~G., and H.~V. {Cane} (2010), {Near-Earth Interplanetary
  Coronal Mass Ejections During Solar Cycle 23 (1996 - 2009): Catalog and
  Summary of Properties}, \textit{Solar Physics}, \textit{264}, 189--237,
  \doi{10.1007/s11207-010-9568-6}.

\bibitem[{\textit{{Rodriguez} et~al.}(2011)\textit{{Rodriguez}, {Mierla},
  {Zhukov}, {West}, and {Kilpua}}}]{Rod11}
{Rodriguez}, L., M.~{Mierla}, A.~N. {Zhukov}, M.~{West}, and E.~{Kilpua}
  (2011), {Linking Remote-Sensing and In Situ Observations of Coronal Mass
  Ejections Using STEREO}, \textit{Solar Physics}, \textit{270}, 561--573,
  \doi{10.1007/s11207-011-9784-8}.

\bibitem[{\textit{{Sachdeva} et~al.}(2017)\textit{{Sachdeva}, {Subramanian},
  {Vourlidas}, and {Bothmer}}}]{Sac17}
{Sachdeva}, N., P.~{Subramanian}, A.~{Vourlidas}, and V.~{Bothmer} (2017), {CME
  dynamics using STEREO and LASCO observations: the relative importance of
  Lorentz forces and solar wind drag}, \textit{ArXiv e-prints}.

\bibitem[{\textit{{Savani} et~al.}(2010)\textit{{Savani}, {Owens}, {Rouillard},
  {Forsyth}, and {Davies}}}]{Sav10}
{Savani}, N.~P., M.~J. {Owens}, A.~P. {Rouillard}, R.~J. {Forsyth}, and J.~A.
  {Davies} (2010), {Observational Evidence of a Coronal Mass Ejection
  Distortion Directly Attributable to a Structured Solar Wind}, \textit{Astrophysical Journal Letters},
  \textit{714}, L128--L132, \doi{10.1088/2041-8205/714/1/L128}.

\bibitem[{\textit{{Savani} et~al.}(2015)\textit{{Savani}, {Vourlidas}, {Szabo},
  {Mays}, {Richardson}, {Thompson}, {Pulkkinen}, {Evans}, and
  {Nieves-Chinchilla}}}]{Sav15}
{Savani}, N.~P., A.~{Vourlidas}, A.~{Szabo}, M.~L. {Mays}, I.~G. {Richardson},
  B.~J. {Thompson}, A.~{Pulkkinen}, R.~{Evans}, and T.~{Nieves-Chinchilla}
  (2015), {Predicting the magnetic vectors within coronal mass ejections
  arriving at Earth: 1. Initial architecture}, \textit{Space Weather},
  \textit{13}, 374--385, \doi{10.1002/2015SW001171}.

\bibitem[{\textit{{Selvakumaran} et~al.}(2016)\textit{{Selvakumaran},
  {Veenadhari}, {Akiyama}, {Pandya}, {Gopalswamy}, {Yashiro}, {Kumar},
  {M{\"a}kel{\"a}}, and {Xie}}}]{Sel16}
{Selvakumaran}, R., B.~{Veenadhari}, S.~{Akiyama}, M.~{Pandya},
  N.~{Gopalswamy}, S.~{Yashiro}, S.~{Kumar}, P.~{M{\"a}kel{\"a}}, and H.~{Xie}
  (2016), {On the reduced geoeffectiveness of solar cycle 24: A moderate storm
  perspective}, \textit{Journal of Geophysical Research (Space Physics)},
  \textit{121}, 8188--8202, \doi{10.1002/2016JA022885}.

\bibitem[{\textit{{Shiota} and {Kataoka}}(2016)}]{Shi16}
{Shiota}, D., and R.~{Kataoka} (2016), {Magnetohydrodynamic simulation of
  interplanetary propagation of multiple coronal mass ejections with internal
  magnetic flux rope (SUSANOO-CME)}, \textit{Space Weather}, \textit{14},
  56--75, \doi{10.1002/2015SW001308}.

\bibitem[{\textit{{Temmer} et~al.}(2017)\textit{{Temmer}, {Thalmann},
  {Dissauer}, {Veronig}, {Tschernitz}, {Hinterreiter}, and
  {Rodriguez}}}]{Tem17}
{Temmer}, M., J.~K. {Thalmann}, K.~{Dissauer}, A.~M. {Veronig},
  J.~{Tschernitz}, J.~{Hinterreiter}, and L.~{Rodriguez} (2017), {On flare-CME
  characteristics from Sun to Earth combining remote-sensing image data with
  in-situ measurements supported by modeling}, \textit{ArXiv e-prints}.

\bibitem[{\textit{{Thernisien} et~al.}(2009)\textit{{Thernisien}, {Vourlidas},
  and {Howard}}}]{The09}
{Thernisien}, A., A.~{Vourlidas}, and R.~A. {Howard} (2009), {Forward Modeling
  of Coronal Mass Ejections Using STEREO/SECCHI Data}, \textit{Solar Physics},
  \textit{256}, 111--130, \doi{10.1007/s11207-009-9346-5}.

\bibitem[{\textit{{Thernisien} et~al.}(2006)\textit{{Thernisien}, {Howard}, and
  {Vourlidas}}}]{The06}
{Thernisien}, A.~F.~R., R.~A. {Howard}, and A.~{Vourlidas} (2006), {Modeling of
  Flux Rope Coronal Mass Ejections}, \textit{Astrophysical Journal }, \textit{652}, 763--773,
  \doi{10.1086/508254}.

\bibitem[{\textit{{Thompson} et~al.}(2012)\textit{{Thompson}, {Kliem}, and
  {T{\"o}r{\"o}k}}}]{Tho12}
{Thompson}, W.~T., B.~{Kliem}, and T.~{T{\"o}r{\"o}k} (2012), {3D
  Reconstruction of a Rotating Erupting Prominence}, \textit{Solar Physics},
  \textit{276}, 241--259, \doi{10.1007/s11207-011-9868-5}.

\bibitem[{\textit{{Vourlidas} et~al.}(2011)\textit{{Vourlidas}, {Colaninno},
  {Nieves-Chinchilla}, and {Stenborg}}}]{Vou11}
{Vourlidas}, A., R.~{Colaninno}, T.~{Nieves-Chinchilla}, and G.~{Stenborg}
  (2011), {The First Observation of a Rapidly Rotating Coronal Mass Ejection in
  the Middle Corona}, \textit{Astrophysical Journal Letters}, \textit{733}, L23,
  \doi{10.1088/2041-8205/733/2/L23}.

\bibitem[{\textit{{Vr{\v s}nak} et~al.}(2007)\textit{{Vr{\v s}nak}, {Sudar},
  {Ru{\v z}djak}, and {{\v Z}ic}}}]{Vrs07b}
{Vr{\v s}nak}, B., D.~{Sudar}, D.~{Ru{\v z}djak}, and T.~{{\v Z}ic} (2007),
  {Projection effects in coronal mass ejections}, \textit{Astronomy and Astrophysics}, \textit{469},
  339--346, \doi{10.1051/0004-6361:20077175}.

\bibitem[{\textit{{Wang} et~al.}(2004)\textit{{Wang}, {Shen}, {Wang}, and
  {Ye}}}]{Wan04}
{Wang}, Y., C.~{Shen}, S.~{Wang}, and P.~{Ye} (2004), {Deflection of coronal
  mass ejection in the interplanetary medium}, \textit{Solar Physics}, \textit{222},
  329--343, \doi{10.1023/B:SOLA.0000043576.21942.aa}.

\bibitem[{\textit{{Wang} et~al.}(2014)\textit{{Wang}, {Wang}, {Shen}, {Shen},
  and {Lugaz}}}]{Wan14}
{Wang}, Y., B.~{Wang}, C.~{Shen}, F.~{Shen}, and N.~{Lugaz} (2014), {Deflected
  propagation of a coronal mass ejection from the corona to interplanetary
  space}, \textit{Journal of Geophysical Research (Space Physics)},
  \textit{119}, 5117--5132, \doi{10.1002/2013JA019537}.

\bibitem[{\textit{{Wood} et~al.}(2017)\textit{{Wood}, {Wu}, {Lepping},
  {Nieves-Chinchilla}, {Howard}, {Linton}, and {Socker}}}]{Woo17}
{Wood}, B.~E., C.-C. {Wu}, R.~P. {Lepping}, T.~{Nieves-Chinchilla}, R.~A.
  {Howard}, M.~G. {Linton}, and D.~G. {Socker} (2017), {A STEREO Survey of
  Magnetic Cloud Coronal Mass Ejections Observed at Earth in 2008-2012},
  \textit{Astrophysical Journal s}, \textit{229}, 29, \doi{10.3847/1538-4365/229/2/29}.

\bibitem[{\textit{{Xie} et~al.}(2013)\textit{{Xie}, {Gopalswamy}, and
  {St.~Cyr}}}]{Xie13}
{Xie}, H., N.~{Gopalswamy}, and O.~C. {St.~Cyr} (2013), {Near-Sun Flux-Rope
  Structure of CMEs}, \textit{Solar Physics}, \textit{284}, 47--58,
  \doi{10.1007/s11207-012-0209-0}.

\bibitem[{\textit{{Xiong} et~al.}(2007)\textit{{Xiong}, {Zheng}, {Wu}, {Wang},
  and {Wang}}}]{Xio07}
{Xiong}, M., H.~{Zheng}, S.~T. {Wu}, Y.~{Wang}, and S.~{Wang} (2007),
  {Magnetohydrodynamic simulation of the interaction between two interplanetary
  magnetic clouds and its consequent geoeffectiveness}, \textit{Journal of
  Geophysical Research (Space Physics)}, \textit{112}, A11103,
  \doi{10.1029/2007JA012320}.

\bibitem[{\textit{{Yashiro} et~al.}(2004)\textit{{Yashiro}, {Gopalswamy},
  {Michalek}, {St.~Cyr}, {Plunkett}, {Rich}, and {Howard}}}]{Yas04}
{Yashiro}, S., N.~{Gopalswamy}, G.~{Michalek}, O.~C. {St.~Cyr}, S.~P.
  {Plunkett}, N.~B. {Rich}, and R.~A. {Howard} (2004), {A catalog of white
  light coronal mass ejections observed by the SOHO spacecraft},
  \textit{Journal of Geophysical Research (Space Physics)}, \textit{109},
  A07105, \doi{10.1029/2003JA010282}.

\bibitem[{\textit{{Yurchyshyn} et~al.}(2005)\textit{{Yurchyshyn}, {Yashiro},
  {Abramenko}, {Wang}, and {Gopalswamy}}}]{Yur05}
{Yurchyshyn}, V., S.~{Yashiro}, V.~{Abramenko}, H.~{Wang}, and N.~{Gopalswamy}
  (2005), {Statistical Distributions of Speeds of Coronal Mass Ejections},
  \textit{Astrophysical Journal }, \textit{619}, 599--603, \doi{10.1086/426129}.

\bibitem[{\textit{{Zhang} and {Dere}}(2006)}]{Zha06}
{Zhang}, J., and K.~P. {Dere} (2006), {A Statistical Study of Main and Residual
  Accelerations of Coronal Mass Ejections}, \textit{Astrophysical Journal }, \textit{649},
  1100--1109, \doi{10.1086/506903}.

\bibitem[{\textit{{Zhou} et~al.}(2014)\textit{{Zhou}, {Feng}, and
  {Zhao}}}]{Zho14}
{Zhou}, Y., X.~{Feng}, and X.~{Zhao} (2014), {Using a 3-D MHD simulation to
  interpret propagation and evolution of a coronal mass ejection observed by
  multiple spacecraft: The 3 April 2010 event}, \textit{Journal of Geophysical
  Research (Space Physics)}, \textit{119}, 9321--9333,
  \doi{10.1002/2014JA020347}.

\end{thebibliography}
\end{document}


\supportinginfo{Sun-to-Earth Evolution of CMEs using Observations and Simulations}

\authors{C. Kay\affil{1} and N. Gopalswamy\affil{1}}
\affiliation{1}{Solar Physics Laboratory, NASA Goddard Space Flight Center, Greenbelt, MD, USA}


\correspondingauthor{C. Kay}{christina.d.kay@nasa.gov}

%
%

\section*{Contents}
\begin{enumerate}
\item  Table \ref{tab:Obs} shows the parameters for the GCS visual fit in both COR1 and COR2

\renewcommand{\baselinestretch}{0.5}
\renewcommand{\arraystretch}{0.7}
\begin{sidewaystable}
\caption{Parameters for the GCS fits to observations in both COR1 and COR2}
\label{tab:Obs}
\begin{tabular}{ >{\raggedleft}p{0.5cm} >{\raggedleft}p{3cm} >{\raggedleft}p{1cm} >{\raggedleft}p{1.1cm} >{\raggedleft}p{1.1cm} >{\raggedleft}p{1.2cm} >{\raggedleft}p{1cm} >{\raggedleft}p{1.1cm} >{\raggedleft}p{1cm} >{\raggedleft}p{1.1cm} >{\raggedleft}p{1.1cm} >{\raggedleft}p{1.2cm} >{\raggedleft}p{1cm} >{\raggedleft\arraybackslash}p{1.1cm} }
 \hline
 \multicolumn{14}{c}{Fits to Observations} \\
 \hline
ID & Date (COR1) & COR1 Lat [$^{\circ}$] & COR1 Lon [$^{\circ}$]& COR1 Tilt [$^{\circ}$]& COR1 R [$R_S$] & COR1 $\kappa$ & COR1 AW [$^{\circ}$]& COR2 Lat [$^{\circ}$]& COR2 Lon [$^{\circ}$]& COR2 Tilt [$^{\circ}$]& COR2 R [$R_S$]& COR2 $\kappa$ & COR2 AW [$^{\circ}$]\\
 \hline
 1 & 15 Nov 2007 23:37$^{*}$ &  ---  & ---   & ---   &  --- & ---   & --- &   -6.2 & 247.1 &  17.3 &  8.36 & 0.281 & 55.1 \\
 2 & 12 Dec 2008 15:07$^{*}$ &  ---  & ---   & ---   &  --- & ---   & --- &   10.6 &  68.2 & -47.0 & 18.21 & 0.136 & 28.5 \\
 3 & 16 Dec 2009 01:45 &  21.8 & 252.7 &  73.8 & 2.21 & 0.130 &  8.1 &   6.7 & 248.2 &  59.2 & 14.00 & 0.111 & 19.6 \\
 4 & 07 Feb 2010 02:55 &   1.7 & 257.1 &  46.4 & 2.43 & 0.158 & 12.9 & -15.1 & 276.1 &  40.8 & 16.07 & 0.222 & 27.0 \\ 
 5 & 03 Apr 2010 09:15 & -25.2 & 258.3 & -63.2 & 2.14 & 0.142 & 14.3 & -25.2 & 258.3 & -63.2 & 12.64 & 0.142 & 30.5 \\
 6 & 08 Apr 2010 03:25 &   8.4 & 191.2 &  49.8 & 2.36 & 0.164 & 20.4 &  -5.6 & 194.5 &  30.2 & 13.57 & 0.260 & 20.4 \\
 7 & 24 May 2010 13:55 &   0.6 & 314.2 & -57.0 & 2.87 & 0.133 & 14.5 &  -2.8 & 307.4 & -68.8 & 14.29 & 0.234 & 31.6 \\
 8 & 01 Aug 2010 08:00 &  22.4 &  78.3 & -56.5 & 2.21 & 0.127 & 21.2 &  24.0 &  79.4 & -52.0 & 13.64 & 0.231 & 52.3 \\
 9 & 15 Feb 2011 02:05 & -16.2 &  35.8 &  -8.4 & 2.93 & 0.247 & 35.8 & -12.9 &  35.8 & -12.9 & 12.07 & 0.270 & 45.0 \\
10 & 03 Mar 2011 04:20 & -13.4 & 186.7 &  35.8 & 2.92 & 0.225 & 35.7 & -19.6 & 186.7 &  24.0 & 14.04 & 0.231 & 38.6 \\
11 & 02 Aug 2011 06:10 &  14.0 & 334.3 & -77.1 & 1.93 & 0.167 & 19.3 &  11.7 & 334.3 & -79.9 & 10.36 & 0.231 & 41.4 \\
12 & 04 Aug 2011 04:00 &  15.7 & 329.8 &  47.0 & 2.79 & 0.185 & 23.5 &  14.0 & 329.8 &  45.8 & 12.57 & 0.646 & 58.7 \\
13 & 06 Sep 2011 01:55 &  22.4 & 244.8 &  28.3 & 2.14 & 0.345 & 28.2 &  25.2 & 249.3 &  20.7 & 14.14 & 0.345 & 33.5 \\
14 & 06 Sep 2011 22:25 &  20.1 & 221.4 &  76.6 & 2.00 & 0.173 &  8.9 &  33.0 & 221.4 &  65.2 & 13.43 & 0.376 & 48.9 \\
15 & 13 Sep 2011 22:45 &  22.9 & 115.2 &  87.7 & 2.57 & 0.133 & 24.3 &  20.7 & 117.4 &  85.5 & 14.57 & 0.148 & 25.4 \\
16 & 19 Sep 2011 06:10 &  36.3 &  47.0 &  83.3 & 2.50 & 0.188 & 19.0 &  40.8 &  47.0 &  85.5 & 12.64 & 0.274 & 56.7 \\
17 & 24 Sep 2011 12:50 &  12.3 & 301.9 & -71.6 & 3.14 & 0.265 & 53.7 &  10.1 & 310.8 & -60.9 & 11.21 & 0.580 & 73.0 \\
18 & 01 Oct 2011 10:05 &   1.1 & 252.6 & -27.4 & 2.07 & 0.314 & 25.4 &   0.6 & 251.6 & -27.4 & 10.57 & 0.379 & 41.1 \\
19 & 09 Nov 2011 13:30 &  25.7 &  59.3 & -51.6 & 3.42 & 0.191 & 25.2 &  25.7 &  59.3 & -52.0 & 13.64 & 0.361 & 57.3 \\
20 & 26 Nov 2011 06:50 &  24.0 & 246.0 & -62.0 & 2.21 & 0.161 & 16.8 &  39.7 & 237.0 & -63.2 & 12.93 & 0.462 & 81.6 \\
21 & 26 Dec 2011 12:50 &  17.9 & 197.6 &  85.0 & 2.07 & 0.161 & 17.6 &  17.9 & 197.6 &  82.7 & 13.71 & 0.231 & 28.5 \\
22 & 18 Jan 2012 10:05 & -25.2 & 234.8 &  53.7 & 2.50 & 0.127 & 12.6 & -29.6 & 241.5 &  55.3 & 13.57 & 0.115 & 27.9 \\
23 & 19 Jan 2012 13:30 &  39.7 & 211.3 & -71.0 & 2.71 & 0.204 & 24.3 &  44.7 & 211.3 & -71.0 & 14.57 & 0.270 & 68.8 \\
24 & 07 Mar 2012 06:50 &  24.0 & 301.9 & -53.7 & 2.36 & 0.216 & 26.8 &  36.9 & 303.0 & -53.7 & 11.07 & 0.373 & 52.8 \\
25 & 13 Mar 2012 11:30 &  19.6 & 294.0 &  49.2 & 2.64 & 0.370 & 30.2 &  19.6 & 294.0 &  49.2 &  6.86 & 0.689 & 60.4 \\
26 & 11 May 2012 23:40 & -15.1 & 159.9 &  53.1 & 2.64 & 0.185 & 17.1 & -16.8 & 159.9 &  53.1 & 13.71 & 0.204 & 42.2 \\
27 & 14 Jun 2012 13:50 & -21.2 &  89.4 &  69.3 & 2.86 & 0.167 & 19.6 & -22.4 &  90.6 &  69.3 & 15.79 & 0.262 & 48.1 \\
28 & 02 Jul 2012 06:35 & -16.2 & 215.8 &  85.5 & 2.64 & 0.142 & 17.9 & -15.1 & 216.9 &  85.5 & 15.43 & 0.213 & 30.2 \\
29 & 12 Jul 2012 16:25 & -15.1 &  81.6 &  58.1 & 2.29 & 0.164 & 22.1 & -15.1 &  81.6 &  58.1 & 15.00 & 0.505 & 58.7 \\
30 & 27 Sep 2012 23:40 &  11.2 & 154.3 & -80.5 & 2.07 & 0.105 & 11.2 &  12.3 & 150.9 & -80.5 & 14.57 & 0.342 & 64.8 \\
31 & 05 Oct 2012 02:15 & -22.9 &  48.1 &  60.4 & 2.93 & 0.155 & 17.3 & -19.0 &  48.1 &  56.5 & 14.71 & 0.191 & 52.0 \\
32 & 27 Oct 2012 15:40 &  10.1 & 115.2 &  62.6 & 2.93 & 0.155 & 17.3 &  10.1 & 115.2 &  62.6 & 13.64 & 0.158 & 24.9 \\
33 & 09 Nov 2012 14:40 & -13.4 & 287.3 &  45.8 & 2.93 & 0.155 & 12.3 & -13.4 & 289.6 &  45.8 & 16.57 & 0.250 & 36.3 \\
34 & 20 Nov 2012 11:30 &   7.3 & 169.9 &  64.9 & 2.14 & 0.164 & 20.1 &  17.3 & 168.8 &  59.8 & 12.21 & 0.250 & 33.8 \\
35 & 23 Nov 2012 13:05 & -25.2 & 100.6 &  49.2 & 2.57 & 0.087 & 14.5 & -14.5 &  98.4 &  49.2 & 15.36 & 0.357 & 56.7 \\
36 & 15 Mar 2013 06:25 &   6.2 &  69.3 & -72.7 & 2.36 & 0.250 &  5.6 &   3.4 &  66.0 & -72.7 & 17.93 & 0.327 & 64.8 \\
37 & 11 Apr 2013 07:10 &   1.1 &  87.2 & -77.7 & 2.64 & 0.250 & 15.7 &   1.1 &  90.6 & -78.3 & 17.50 & 0.287 & 43.6 \\
38 & 09 Jul 2013 14:45 &  10.6 & 343.2 & -62.1 & 3.00 & 0.311 & 17.9 &   6.7 & 343.2 & -62.6 & 15.71 & 0.299 & 53.7 \\
39 & 29 Sep 2013 21:55 &  19.6 & 355.6 &  82.7 & 2.71 & 0.311 & 12.3 &  21.2 & 355.6 &  82.7 & 15.36 & 0.514 & 66.8 \\
40 & 06 Oct 2013 14:15 &   0.6 & 250.4 & -78.3 & 3.93 & 0.096 &  8.4 &   2.8 & 250.4 & -78.3 & 18.43 & 0.431 & 34.6 \\
41 & 12 Dec 2013 03:25 & -25.7 & 136.4 & -38.6 & 2.07 & 0.388 & 34.4 & -31.9 & 136.4 & -30.8 & 14.00 & 0.527 & 75.5 \\
42 & 04 Feb 2014 02:54$^{*}$& ---   & ---   &  ---  & ---  &  ---  & ---  & -29.6 & 118.5 & -64.3 & 12.57 & 0.382 & 51.7 \\
43 & 12 Feb 2014 05:15 &  -6.7 & 353.3 &  74.9 & 3.57 & 0.185 & 10.6 &  -3.9 & 351.1 &  74.9 & 20.21 & 0.327 & 38.9 \\
44 & 18 Apr 2014 13:00 & -20.7 & 229.2 &  79.9 & 3.57 & 0.136 &  9.2 & -23.5 & 227.0 &  79.4 & 15.64 & 0.474 & 71.0 \\
45 & 04 Jun 2014 16:15 & -39.1 & 286.2 &  55.3 & 3.93 & 0.179 & 14.8 & -41.4 & 286.2 &  56.5 & 14.71 & 0.321 & 46.1 \\
 
 \hline
\end{tabular}
\end{sidewaystable}

\item Table \ref{tab:mod} shows the free parameters for the ForeCAT and FIDO simulation and the score for the FIDO fits.

\begin{sidewaystable}
\caption{Parameters for the ForeCAT and FIDO models and the resulting score}
\label{tab:mod}
\begin{tabular}{>{\raggedleft}p{0.5cm} >{\raggedleft}p{3cm} >{\raggedleft}p{0.75cm} >{\raggedleft}p{0.8cm} >{\raggedleft}p{0.75cm} >{\raggedleft}p{0.45cm} >{\raggedleft}p{0.6cm} >{\raggedleft}p{0.75cm} >{\raggedleft}p{0.8cm} >{\raggedleft}p{1.2cm} >{\raggedleft}p{0.75cm} >{\raggedleft}p{0.8cm} >{\raggedleft}p{0.75cm} >{\raggedleft}p{0.65cm} >{\raggedleft}p{0.75cm} >{\raggedleft}p{0.65cm} >{\raggedleft}p{0.9cm} >{\raggedleft}p{0.7cm} >{\raggedleft\arraybackslash}p{.7cm}}
 \hline
 \multicolumn{18}{c}{Model Parameters} \\
 \hline
ID & Date (1 AU) & Lat$_i$ [$^{\circ}$] & Lon$_i$ [$^{\circ}$] & Tilt$_i$ [$^{\circ}$] & AW [$^{\circ}$] &A  & B$_{cor}$ & v$_{cor}$ [km/s] &  M $\;$ [10$^{15}$ g] & Lat$_f$ [$^{\circ}$] & Lon$_f$ [$^{\circ}$] & Tilt$_f$ [$^{\circ}$]  & B$_{1 AU}$ & v$_{1 AU}$ [km/s] & Pol. & Qual. & $\delta_{FIDO}$ & $\delta_{GCS}$ \\
\hline
 1 & 20 Nov 2007 00:33 &  -9.8 & 241.0 &  22.0 & 55 & 0.8 & 0.20 &  350 &  0.38 &  -9.2 & 250.3 &  21.0 & 0.64 & 460 & +$-$ & MC 1& 0.70 & 0.84 \\ 
 2 & 17 Dec 2008 03:35 &  47.0 &  82.2 & -28.0 & 25 & 0.8 & 0.20 &  480 &  1.00 &  12.2 &  66.1 & -51.6 & 0.11 & 350 & +$-$ & MC 1& 0.73 & 0.76 \\ 
 3 & 19 Dec 2009 13:00 &  22.9 & 254.3 &  76.5 & 20 & 0.7 & 0.20 &  360 &  1.50 &   3.8 & 252.7 &  29.7 & 0.50 & 380 & ++ & 3 & 0.91 & 1.11 \\ 
 4 & 11 Feb 2010 08:00 &  22.6 & 243.8 &  65.0 & 25 & 0.8 & 0.10 &  250 &  1.00 &  -8.1 & 269.5 &  35.2 & 0.28 & 360 & +$-$ & 3 & 0.67 & 1.37 \\  
 5 & 05 Apr 2010 12:00 & -24.1 & 261.0 & -74.0 & 30 & 0.7 & 0.10 &  660 &  3.00 & -25.4 & 260.4 & -73.2 & 0.37 & 640 & $-$+ & MC 2& 0.85 & 0.93 \\ 
 6 & 12 Apr 2010 01:00 &  23.3 & 178.7 &  86.0 & 20 & 0.8 & 0.18 &  250 &  1.30 &  -3.9 & 198.6 &  33.6 & 0.27 & 410 & ++ & MC 2 & 0.58 & 0.88 \\ 
 7 & 28 May 2010 19:12 &  15.0 & 321.8 & -55.0 & 30 & 0.7 & 0.20 &  475 &  1.00 &  -4.6 & 306.8 & -66.7 & 0.26 & 360 & +$-$ & MC 1& 0.33 & 0.51 \\ 
 8 & 04 Aug 2010 10:00 &  20.6 &  77.6 & -55.0 & 44 & 0.8 & 0.10 & 1300 &  5.00 &  26.5 &  78.4 & -49.6 & 0.54 & 530 & $-$$-$ & MC 2 & 0.57 & 0.55 \\ 
 9 & 18 Feb 2011 19:55 & -19.0 &  33.5 &   5.0 & 45 & 1.0 & 0.11 & 1600 &  4.30 & -15.4 &  33.5 &  -6.9 & 0.44 & 470 & ++   & 2 & 0.62  & 0.73 \\ 
10 & 06 Mar 2011 09:00 & -12.2 & 184.9 &  37.0 & 40 & 0.9 & 0.15 &  380 &  1.40 & -13.1 & 183.8 &  34.5 & 0.73 & 430 & +$-$ & 3 & 0.89 & 1.05 \\ 
11 & 05 Aug 2011 05:00 &  16.9 & 334.0 & -75.0 & 40 & 0.7 & 0.20 &  600 &  5.10 &   8.7 & 334.1 & -79.8 & 0.48 & 430 & $-$+ & 3 & 0.51 & 0.68 \\ 
12 & 06 Aug 2011 22:00 &  16.1 & 325.6 &  42.0 & 55 & 0.7 & 0.15 & 1200 & 11.00 &  11.6 & 328.8 &  44.6 & 0.54 & 540 & ++ & 2 & 0.54 & 0.57 \\ 
13 & 08 Sep 2011 10:00 &  18.0 & 236.5 &  22.0 & 35 & 0.8 & 0.10 & 1000 &  2.30 &  23.6 & 251.7 &  27.6 & 0.12 & 320 & $-$+ & 3 & 0.99 & 1.15 \\ 
14 & 10 Sep 2011 03:00 &  15.9 & 222.6 &  79.0 & 45 & 0.7 & 0.10 &  580 & 15.00 &  31.1 & 219.3 &  68.3 & 0.11 & 470 & ++ & MC 2 & 0.94 & 0.93 \\  
15 & 17 Sep 2011 15:35 &  23.4 & 121.7 & -50.0 & 25 & 0.7 & 0.10 &  460 &  2.50 &  23.0 & 116.0 &  88.1 & 0.42 & 430 & ++ & MC 2 & 0.98 & 0.95 \\  
16 & 22 Sep 2011 15:00 &  25.0 &  50.5 & -33.0 & 45 & 0.7 & 0.15 &  640 & 10.00 &  37.7 &  42.1 & -78.8 & 0.09 & 390 & $-$+ & 3 & 0.45 & 0.93 \\  
17 & 26 Sep 2011 19:45 &  12.5 & 290.6 &  81.0 & 70 & 1.0 & 0.20 & 2100 & 16.00 &  12.8 & 307.6 & -60.9 & 0.64 & 580 & +$-$ & 3 & 1.05 & 1.06 \\  
18 & 05 Oct 2011 10:00 &   9.0 & 251.8 & -25.0 & 35 & 0.7 & 0.15 &  450 &  3.70 &   3.1 & 250.0 & -22.0 & 0.54 & 450 & $-$+ & 2 & 0.97 & 0.95 \\  
19 & 13 Nov 2011 10:00 &  21.5 &  60.5 & -53.0 & 55 & 0.7 & 0.10 &  900 & 14.00 &  23.5 &  57.6 & -54.3 & 0.59 & 370 & +$-$ & 3 & 0.74 & 0.78 \\  
20 & 29 Nov 2011 00:00 &  22.8 & 262.0 & -49.5 & 80 & 0.8 & 0.20 & 1000 & 12.00 &  48.7 & 231.1 & -62.5 & 0.34 & 450 & ++ & MC 2 & 1.01 & 1.06 \\  
21 & 29 Dec 2011 00:00 &  12.9 & 198.0 &  88.1 & 25 & 0.8 & 0.10 &  740 &  4.30 &  16.6 & 196.5 &  88.6 & 0.45 & 400 & $-$+ & 3 & 0.33 & 0.96 \\  
22 & 21 Jan 2012 06:00 & -22.8 & 225.2 &  43.0 & 26 & 0.8 & 0.20 &  270 &  2.20 & -27.4 & 243.2 &  58.9 & 0.32 & 320 & ++ & MC 3& 0.84 & 0.86 \\  
23 & 22 Jan 2012 11:24 &  29.3 & 209.7 & -74.0 & 50 & 0.7 & 0.20 & 1500 & 19.00 &  38.1 & 211.7 & -71.4 & 0.39 & 450 & ++ & MC 2& 0.87 & 0.77 \\  
24 & 09 Mar 2012 03:00 &  18.4 & 298.5 & -54.0 & 50 & 0.8 & 0.20 & 2600 & 14.00 &  30.9 & 303.3 & -54.5 & 0.38 & 550 & $-$$-$ & 2 & 0.84 & 0.91 \\  
25 & 15 Mar 2012 17:00 &  18.8 & 303.5 &  44.0 & 60 & 0.9 & 0.20 & 1990 & 23.00 &  16.7 & 301.1 &  46.9 & 0.63 & 680 & +$-$ & 2 & 0.72 & 0.89 \\  
26 & 16 May 2012 16:00 & -14.1 & 168.2 &  42.0 & 40 & 0.8 & 0.15 &  800 &  4.60 & -14.2 & 165.9 &  42.0 & 0.26 & 370 & +$-$ & MC 2 & 0.95 & 0.97 \\  
27 & 16 Jun 2012 23:00  & -13.6 &  87.3 &  60.0 & 45 & 0.7 & 0.10 &  980 & 12.00 & -17.1 &  89.8 &  67.5 & 0.11 & 440 & $-$+ & MC 1 & 0.87 & 1.01\\  
28 & 05 Jul 2012 00:00 & -17.1 & 204.5 & -63.0 & 30 & 0.8 & 0.20 & 1080 &  6.00 & -16.1 & 217.4 &  80.2 & 0.29 & 470 & +$-$ & 3 & 0.91 & 0.99 \\  
29 & 15 Jul 2012 06:14 & -13.3 &  84.5 &  50.0 & 46 & 0.7 & 0.10 & 2260 &  6.90 & -15.2 &  84.1 &  51.6 & 0.56 & 490 & $-$+ & MC 1 & 0.54 & 0.60 \\  
30 & 01 Oct 2012 00:00 &   8.5 & 171.8 & -80.0 & 50 & 0.8 & 0.20 &  950 &  9.20 &  15.4 & 147.0 & -75.6 & 0.24 & 370 & $-$$-$ & MC 2 & 0.67 & 0.86 \\  
31 & 08 Oct 2012 18:00 & -23.3 &  48.5 &  58.0 & 50 & 0.8 & 0.10 &  800 & 10.00 & -22.2 &  45.2 &  60.5 & 0.27 & 390 & $-$+ & MC 2 & 0.48 & 0.48 \\  
32 & 01 Nov 2012 00:00 &  11.7 & 128.5 &  83.0 & 25 & 0.8 & 0.20 &  510 &  4.90 &   7.8 & 116.3 &  61.8 & 0.36 & 340 & $-$+ & MC 1 & 0.43 & 0.56 \\  
33 & 13 Nov 2012 08:23 & -16.4 & 284.2 &  55.0 & 30 & 0.8 & 0.20 &  600 &  5.00 & -14.8 & 291.1 &  62.1 & 0.16 & 380 & $-$$-$ & MC 2 & 0.93 & 1.12 \\  
34 & 24 Nov 2012 12:00 &  15.2 & 171.7 &  87.0 & 30 & 0.8 & 0.20 &  650 &  8.40 &  16.6 & 167.9 &  69.0 & 0.24 & 380 & +$-$ & 2 & 0.71 & 0.80\\  
35 & 26 Nov 2012 12:00 & -29.2 & 120.8 &  47.0 & 55 & 0.7 & 0.10 &  500 &  6.90 &  -9.9 & 103.1 &  56.2 & 0.34 & 450 & ++ & 3 & 0.71 & 0.79 \\  
36 & 17 Mar 2013 15:00 &  11.2 &  73.0 & -73.4 & 55 & 0.8 & 0.20 & 1250 & 13.00 &   1.6 &  66.2 & -70.5 & 0.63 & 520 & $-$+ & MC 3 & 0.82 & 0.81 \\  
37 & 14 Apr 2013 17:00 &  10.1 &  83.9 & -50.0 & 40 & 0.8 & 0.20 &  860 & 22.00 &   6.0 &  93.4 & -75.4 & 0.39 & 410 & $-$$-$ & MC 1 & 0.24 & 0.38 \\  
38 & 13 Jul 2013 05:00 &  14.5 & 341.0 & -36.0 & 50 & 0.7 & 0.10 &  450 &  3.40 &   7.8 & 338.9 & -49.5 & 0.36 & 430 & +$-$ & MC 2 & 0.61 & 0.72 \\  
39 & 02 Oct 2013 23:00 &  12.2 &   9.7 & -50.0 & 60 & 0.9 & 0.20 & 1180 & 22.00 &  20.6 & 353.1 &  75.7 & 0.30 & 470 & $-$+ & MC 2 & 0.47 & 0.82 \\  
40 & 09 Oct 2013 09:00 & -10.8 & 255.3 &  84.0 & 30 & 0.8 & 0.20 &  880 &  0.46 &   4.6 & 243.6 & -69.0 & 0.49 & 480 & ++ & 2 & 0.30 & 0.57 \\  
41 & 15 Dec 2013 16:47 & -24.6 & 136.5 &  63.0 & 75 & 0.8 & 0.20 & 1000 & 14.00 & -27.4 & 136.4 & -27.2 & 0.07 & 460 & ++ & MC 1 & 0.58 & 0.58 \\  
42 & 08 Feb 2014 01:00 & -10.2 & 115.3 & -53.0 & 45 & 0.7 & 0.10 &  530 &  6.90 & -21.1 & 115.6 & -59.1 & 0.28 & 420 & $-$$-$ & MC 3 & 1.0 & 1.02 \\  
43 & 16 Feb 2014 00:42 &  -8.0 & 356.1 &  66.0 & 35 & 0.7 & 0.10 &  490 &  5.00 &  -5.4 & 358.9 &  69.8 & 0.14 & 380 & +$-$ & MC 2 & 0.63 & 1.42 \\  
44 & 21 Apr 2014 07:41 & -14.0 & 247.5 & -70.0 & 70 & 0.8 & 0.20 & 1280 & 20.00 & -22.2 & 224.6 &  78.2 & 0.29 & 540 & +$-$ & 2 & 1.03 & 1.07 \\  
45 & 08 Jun 2014 20:00 & -21.1 & 273.6 &  66.0 & 45 & 0.7 & 0.20 & 1035 & 16.00 & -37.5 & 286.3 &  52.6 & 0.48 & 480 & ++ & 3 & 0.69 & 0.73 \\  
 \hline
\end{tabular}
\end{sidewaystable}

\begin{sidewaystable}
\caption{FIDO Parameters from the Random Walk Best Fits}
\label{tab:BFs}
\begin{tabular}{>{\raggedleft}p{0.5cm} >{\raggedleft}p{0.75cm} >{\raggedleft}p{0.8cm} >{\raggedleft}p{0.75cm} >{\raggedleft}p{0.7cm} >{\raggedleft}p{0.6cm} >{\raggedleft}p{0.75cm} >{\raggedleft}p{0.6cm} >{\raggedleft}p{0.6cm} >{\raggedleft}p{0.6cm} >{\raggedleft}p{0.7cm}  >{\raggedleft}p{0.75cm} >{\raggedleft}p{0.8cm} >{\raggedleft}p{0.75cm} >{\raggedleft}p{0.7cm} >{\raggedleft}p{0.6cm} >{\raggedleft}p{0.75cm}   >{\raggedleft}p{0.6cm} >{\raggedleft}p{0.6cm} >{\raggedleft}p{0.6cm}>{\raggedleft\arraybackslash}p{.7cm}}
 \hline
 \multicolumn{21}{c}{Model Parameters} \\
 \hline
ID & Lat$_{FF}$ [$^{\circ}$] & Lon$_{FF}$ [$^{\circ}$] & Tilt$_{FF}$ [$^{\circ}$] & AW$_{FF}$ [$^{\circ}$] &A$_{FF}$  & B$_{FF}$ & $\delta_{x,FF}$ & $\delta_{y,FF}$ & $\delta_{z,FF}$ & $\delta_{FF}$ & Lat$_{cir}$ [$^{\circ}$] & Lon$_{cir}$ [$^{\circ}$] & Tilt$_{cir}$ [$^{\circ}$] & AW$_{cir}$ [$^{\circ}$] &A$_{cir}$  & B$_{cir}$ &  $\delta_{x,cir}$ & $\delta_{y,cir}$ & $\delta_{z,cir}$ & $\delta_{cir}$  \\
\hline
1  &  -9.3 & 251.1 &  20.9 & 54.4 & 0.80 & 0.64 & 0.39 & 0.16 & 0.39 & 0.61 &  -9.4 & 251.1 &  20.6 & 54.4 & 0.80 & 0.64 & 0.17 & 0.24 & 0.41 & 0.52 \\
2  &  11.9 &  66.3 & -50.7 & 25.6 & 0.80 & 0.11 & 0.30 & 0.40 & 0.40 & 0.68 &  10.9 &  67.0 & -49.8 & 26.5 & 0.77 & 0.13 & 0.33 & 0.30 & 0.45 & 0.67 \\
3  &   8.3 & 254.1 &  26.0 & 25.0 & 0.69 & 0.51 & 0.30 & 0.29 & 0.52 & 0.72 &   6.1 & 251.2 &  24.6 & 25.9 & 0.66 & 0.52 & 0.35 & 0.27 & 0.48 & 0.72 \\
4  &  -4.3 & 274.5 &  33.5 & 25.1 & 0.66 & 0.28 & 0.14 & 0.33 & 0.28 & 0.51 &  -7.9 & 269.7 &  36.1 & 24.3 & 0.74 & 0.29 & 0.16 & 0.22 & 0.26 & 0.42 \\
5  & -26.7 & 262.1 & -74.0 & 27.8 & 0.70 & 0.37 & 0.31 & 0.28 & 0.20 & 0.51 & -26.8 & 261.6 & -74.1 & 27.9 & 0.71 & 0.38 & 0.34 & 0.23 & 0.20 & 0.50 \\
6  &  -5.0 & 199.9 &  33.5 & 20.3 & 0.69 & 0.27 & 0.09 & 0.25 & 0.31 & 0.44 &  -2.7 & 201.6 &  32.6 & 19.4 & 0.73 & 0.27 & 0.08 & 0.08 & 0.12 & 0.19 \\
7  &  -0.1 & 304.8 & -66.0 & 31.2 & 0.78 & 0.26 & 0.17 & 0.09 & 0.12 & 0.25 &  -0.3 & 305.0 & -66.1 & 30.6 & 0.80 & 0.27 & 0.18 & 0.08 & 0.10 & 0.25 \\
8  &  25.2 &  79.9 & -52.2 & 44.9 & 0.79 & 0.54 & 0.30 & 0.28 & 0.20 & 0.48 &  25.6 &  80.0 & -52.4 & 44.5 & 0.81 & 0.54 & 0.33 & 0.28 & 0.19 & 0.49 \\
9  & -16.1 &  33.1 &  -9.3 & 45.9 & 0.69 & 0.45 & 0.17 & 0.30 & 0.41 & 0.59 & -17.0 &  33.7 & -10.2 & 46.0 & 0.69 & 0.46 & 0.14 & 0.32 & 0.43 & 0.62 \\
10 &  -9.7 & 185.3 &  34.4 & 35.7 & 0.96 & 0.71 & 0.51 & 0.28 & 0.24 & 0.68 &  -8.6 & 185.5 &  25.8 & 33.9 & 0.63 & 0.63 & 0.28 & 0.26 & 0.24 & 0.52 \\
11 &   6.6 & 334.5 & -79.9 & 40.0 & 0.71 & 0.48 & 0.13 & 0.18 & 0.31 & 0.40 &   8.7 & 334.1 & -79.9 & 39.9 & 0.70 & 0.48 & 0.10 & 0.20 & 0.05 & 0.24 \\
12 &  10.7 & 329.7 &  42.9 & 55.3 & 0.72 & 0.56 & 0.13 & 0.33 & 0.33 & 0.52 &  12.5 & 330.5 &  43.5 & 53.9 & 0.76 & 0.55 & 0.16 & 0.34 & 0.34 & 0.54 \\
13 &  24.3 & 251.5 &  26.8 & 34.8 & 0.80 & 0.12 & 0.73 & 0.27 & 0.26 & 0.89 &  24.6 & 251.7 &  26.8 & 34.9 & 0.80 & 0.12 & 0.72 & 0.31 & 0.27 & 0.90 \\
14 &  29.0 & 219.6 &  65.7 & 45.5 & 0.63 & 0.12 & 0.47 & 0.25 & 0.56 & 0.90 &  30.0 & 220.6 &  67.1 & 45.5 & 0.67 & 0.10 & 0.40 & 0.15 & 0.58 & 0.79 \\
15 &  22.4 & 115.4 &  87.7 & 25.6 & 0.70 & 0.42 & 0.49 & 0.49 & 0.47 & 0.91 &  22.6 & 115.2 &  87.5 & 25.5 & 0.70 & 0.42 & 0.78 & 0.53 & 0.46 & 1.10 \\
16 &  39.0 &  40.4 & -76.6 & 44.3 & 0.72 & 0.09 & 0.16 & 0.27 & 0.20 & 0.42 &  38.0 &  37.1 & -70.5 & 43.1 & 0.74 & 0.09 & 0.13 & 0.23 & 0.16 & 0.37 \\
17 &  11.1 & 307.9 & -62.0 & 71.0 & 0.81 & 0.65 & 0.72 & 0.45 & 0.38 & 1.04 &  15.0 & 305.6 & -60.2 & 71.7 & 0.96 & 0.67 & 0.73 & 0.44 & 0.37 & 1.05 \\
18 &   3.0 & 247.4 & -26.7 & 35.9 & 0.68 & 0.55 & 0.30 & 0.56 & 0.50 & 0.92 &   0.0 & 249.8 & -30.9 & 37.4 & 0.71 & 0.56 & 0.34 & 0.56 & 0.38 & 0.86 \\
19 &  23.7 &  57.2 & -53.4 & 55.2 & 0.86 & 0.60 & 0.26 & 0.44 & 0.30 & 0.68 &  23.6 &  57.4 & -54.5 & 54.7 & 0.80 & 0.60 & 0.20 & 0.31 & 0.19 & 0.47 \\
20 &  48.4 & 230.7 & -62.2 & 79.8 & 0.81 & 0.34 & 0.36 & 0.51 & 0.60 & 1.01 &  49.3 & 230.4 & -62.4 & 79.8 & 0.83 & 0.34 & 0.45 & 0.58 & 0.60 & 1.04 \\
21 &  16.6 & 195.2 &  89.3 & 25.0 & 0.75 & 0.45 & 0.07 & 0.11 & 0.18 & 0.23 &  16.4 & 196.2 &  88.8 & 25.0 & 0.75 & 0.46 & 0.14 & 0.12 & 0.26 & 0.35 \\
22 & -27.3 & 244.8 &  58.1 & 26.9 & 0.80 & 0.32 & 0.36 & 0.32 & 0.53 & 0.79 & -27.1 & 244.8 &  57.8 & 26.8 & 0.81 & 0.32 & 0.45 & 0.39 & 0.53 & 0.87 \\
23 &  36.4 & 211.9 & -74.6 & 52.5 & 0.69 & 0.38 & 0.31 & 0.39 & 0.35 & 0.70 &  35.5 & 212.7 & -76.2 & 51.7 & 0.74 & 0.39 & 0.19 & 0.39 & 0.27 & 0.62 \\
24 &  28.2 & 307.6 & -53.0 & 49.1 & 0.68 & 0.34 & 0.38 & 0.21 & 0.36 & 0.65 &  30.8 & 307.4 & -52.1 & 48.2 & 0.71 & 0.38 & 0.21 & 0.20 & 0.34 & 0.50 \\
25 &  17.1 & 302.2 &  45.3 & 59.7 & 0.89 & 0.64 & 0.34 & 0.53 & 0.24 & 0.70 &  17.0 & 302.4 &  45.6 & 60.2 & 0.90 & 0.64 & 0.35 & 0.54 & 0.24 & 0.71 \\
26 & -18.5 & 162.5 &  43.5 & 42.1 & 0.48 & 0.25 & 0.17 & 0.40 & 0.75 & 0.90 & -11.2 & 166.0 &  41.3 & 44.4 & 0.50 & 0.28 & 0.26 & 0.32 & 0.62 & 0.82 \\
27 & -17.2 &  89.4 &  67.2 & 45.0 & 0.71 & 0.10 & 0.24 & 0.28 & 0.47 & 0.63 & -17.3 &  89.2 &  66.8 & 45.1 & 0.71 & 0.10 & 0.26 & 0.32 & 0.54 & 0.73 \\
28 & -18.2 & 216.5 &  80.8 & 29.4 & 0.88 & 0.31 & 0.45 & 0.53 & 0.37 & 0.88 & -11.1 & 221.2 &  82.8 & 37.0 & 0.65 & 0.36 & 0.29 & 0.55 & 0.38 & 0.82 \\
29 & -12.1 &  88.8 &  51.0 & 49.0 & 0.61 & 0.57 & 0.26 & 0.17 & 0.19 & 0.39 & -16.5 &  86.5 &  50.9 & 49.5 & 0.57 & 0.57 & 0.19 & 0.18 & 0.19 & 0.35 \\
30 &  15.6 & 145.4 & -74.3 & 49.2 & 0.86 & 0.23 & 0.09 & 0.50 & 0.22 & 0.60 &  14.7 & 146.9 & -73.8 & 50.0 & 0.79 & 0.24 & 0.08 & 0.44 & 0.30 & 0.61 \\
31 & -20.9 &  46.0 &  61.2 & 50.3 & 0.78 & 0.27 & 0.24 & 0.30 & 0.16 & 0.45 & -20.2 &  46.4 &  60.8 & 51.0 & 0.76 & 0.27 & 0.24 & 0.29 & 0.19 & 0.47 \\
32 &   6.3 & 116.4 &  63.4 & 25.1 & 0.80 & 0.37 & 0.20 & 0.26 & 0.15 & 0.38 &   5.9 & 116.2 &  63.6 & 25.4 & 0.80 & 0.37 & 0.18 & 0.19 & 0.14 & 0.33 \\
33 & -15.1 & 290.4 &  59.9 & 29.9 & 0.78 & 0.15 & 0.25 & 0.33 & 0.75 & 0.91 & -14.7 & 290.7 &  61.5 & 30.4 & 0.79 & 0.16 & 0.38 & 0.31 & 0.73 & 0.95 \\
34 &  19.9 & 172.8 &  67.7 & 28.7 & 0.83 & 0.25 & 0.13 & 0.27 & 0.29 & 0.46 &  18.1 & 171.1 &  67.9 & 30.0 & 0.83 & 0.26 & 0.16 & 0.35 & 0.31 & 0.55 \\
35 &  -7.9 & 101.6 &  59.9 & 58.0 & 0.72 & 0.37 & 0.14 & 0.47 & 0.32 & 0.64 & -12.1 &  99.6 &  58.7 & 56.7 & 0.82 & 0.38 & 0.16 & 0.45 & 0.27 & 0.61 \\
36 &   0.4 &  65.8 & -73.4 & 56.3 & 0.73 & 0.63 & 0.35 & 0.29 & 0.42 & 0.79 &  -2.6 &  64.9 & -75.1 & 57.6 & 0.63 & 0.63 & 0.34 & 0.30 & 0.43 & 0.78 \\
37 &   5.6 &  93.5 & -75.4 & 39.9 & 0.81 & 0.39 & 0.14 & 0.11 & 0.09 & 0.24 &   1.0 &  94.2 & -78.2 & 39.9 & 0.83 & 0.39 & 0.16 & 0.12 & 0.11 & 0.25 \\
38 &   8.9 & 340.9 & -44.6 & 48.7 & 0.70 & 0.35 & 0.11 & 0.23 & 0.36 & 0.48 &   8.7 & 340.6 & -42.8 & 49.0 & 0.69 & 0.34 & 0.13 & 0.25 & 0.31 & 0.46 \\
39 &  22.1 & 353.2 &  76.7 & 59.6 & 0.93 & 0.31 & 0.24 & 0.13 & 0.27 & 0.43 &  23.4 & 354.1 &  76.3 & 60.2 & 0.93 & 0.32 & 0.31 & 0.13 & 0.29 & 0.49 \\
40 &   4.4 & 243.7 & -69.2 & 30.0 & 0.80 & 0.49 & 0.15 & 0.17 & 0.15 & 0.29 &   2.4 & 244.5 & -69.5 & 29.9 & 0.80 & 0.49 & 0.14 & 0.19 & 0.14 & 0.29 \\
41 & -27.1 & 137.5 & -27.8 & 74.7 & 0.81 & 0.07 & 0.16 & 0.25 & 0.22 & 0.46 & -26.2 & 137.8 & -28.2 & 75.0 & 0.78 & 0.08 & 0.09 & 0.22 & 0.17 & 0.33 \\
42 & -21.1 & 115.5 & -59.1 & 45.0 & 0.70 & 0.29 & 0.66 & 0.34 & 0.43 & 0.98 & -20.9 & 115.5 & -58.6 & 45.7 & 0.68 & 0.29 & 0.65 & 0.35 & 0.38 & 1.01 \\
43 &  -4.9 & 358.6 &  70.0 & 35.2 & 0.70 & 0.13 & 0.15 & 0.42 & 0.30 & 0.60 &  -5.0 & 358.6 &  70.9 & 35.4 & 0.72 & 0.14 & 0.14 & 0.41 & 0.32 & 0.63 \\
44 & -21.0 & 222.2 &  79.5 & 68.6 & 0.87 & 0.27 & 0.64 & 0.26 & 0.57 & 0.97 & -22.4 & 223.2 &  80.0 & 69.2 & 0.84 & 0.28 & 0.66 & 0.28 & 0.53 & 0.99 \\
45 & -37.3 & 286.0 &  52.5 & 45.5 & 0.60 & 0.46 & 0.42 & 0.35 & 0.26 & 0.78 & -36.8 & 287.8 &  52.9 & 44.1 & 0.65 & 0.47 & 0.42 & 0.32 & 0.25 & 0.63 \\

 \hline
\end{tabular}
\end{sidewaystable}

\end{enumerate}